\documentclass[10pt,a4paper]{article}             

\usepackage{amsmath,amsthm,amssymb,mathrsfs,bm,enumerate,physics}
\usepackage{tikz}
\usetikzlibrary{intersections,calc,arrows.meta}
\usepackage{url}
\usepackage{hyperref}
\usepackage{cleveref}
\usepackage{comment}
\newcommand\myshade{75}
\colorlet{mylinkcolor}{red}
\colorlet{mycitecolor}{green}
\colorlet{myurlcolor}{blue}
\hypersetup{colorlinks=true, anchorcolor=cyan, 
linkcolor=mylinkcolor!\myshade!black, 
citecolor=mycitecolor!\myshade!black,
urlcolor=myurlcolor!\myshade!black } 

\newcommand{\beq}{\begin{equation}}
\newcommand{\eeq}{\end{equation}}
\newcommand{\nn}{\nonumber}
\newcommand{\ra}{\rightarrow}
\newcommand{\lra}{\leftrightarrow}
\newcommand{\qqi}{q-q^{-1}}
\newcommand{\eps}{\epsilon}
\newcommand{\pqCom}[4]{ [\,{#1} \,, {#2}\,]_{({#3},{#4})}    }
\newcommand{\qCom}[3]{ [\,{#1} \,, {#2}\,]_{({#3})}    }

\newcommand{\VT}[2]{ T_{#1}^{({#2})}}

\newcommand{\Op}[3]{ \hat{#1}_{#2}^{({#3})}}
\newcommand{\Lp}[3]{ L_{#1}^{#3({#2})}  }
\newcommand{\Wp}[3]{ W_{#1}^{#3({#2})}  }
\newcommand{\Exp}[1]{\exp\left({#1}\right)}

\theoremstyle{definition}
\newtheorem*{rmk*}{$\ddagger$  Remark}

\theoremstyle{remark}
\newtheorem*{exm*}{$\spadesuit$ Example}

\makeatletter
\long\def\@makefntext#1{\parindent 1em\noindent
\@hangfrom{\hbox to 1.8em{\hss$^{\@thefnmark}$}}#1}
\makeatother
\begin{document}
\topmargin 0pt
\oddsidemargin 0mm
\renewcommand{\thefootnote}{\fnsymbol{footnote}}

\vspace*{0.5cm}

\begin{center}
{\Large Moyal product and Generalized Hom-Lie-Virasoro symmetries \\
in Bloch electron systems}
\vspace{1.5cm}

{\large Haru-Tada Sato${\,}^{a,b}$
\footnote{\,\,\, Corresponding author.  E-mail address: satoh@isfactory.co.jp}
}\\

{${}^{a}$\em Department of Physics, Graduate School of Science, \\Osaka Metropolitan University \\
Nakamozu Campus, Sakai, Osaka 599-8531, Japan}\\

{${}^{b}$\em Department of Data Science, i's Factory Corporation, Ltd.\\
     Kanda-nishiki-cho 2-7-6, Tokyo 101-0054, Japan}\\
%
\end{center}

\vspace{0.1cm}

\abstract{
We explore two variations of the Curtright-Zachos (CZ) deformation of the Virasoro algebra. Firstly, we introduce a scaled CZ algebra that inherits the scaling structure found in the differential operator representation of the magnetic translation (MT) operators. 
We then linearly decompose the scaled CZ generators to derive two types of Hom-Lie deformations of the $W_\infty$ algebra. We discuss $\ast$-bracket formulations of these algebras and their connection to the Moyal product. 
We show that the $\ast$-bracket form of the scaled CZ algebra arises from the Moyal product, while we obtain the second type of deformed $W_\infty$ through a coordinate transformation of the first type of Moyal operators. From a physical point of view, we construct the Hamiltonian of a tight binding model (TBM) using the Wyle matrix representation of the scaled CZ algebra. We note that the integer powers of $q$ are linked to the quantum fluctuations that are inherent in the Moyal product.
}
\vspace{0.5cm}

\begin{description}
\item[Keywords:] quantum plane, $q$-deformation, Virasoro algebra, Hom-Lie algebra, magnetic translation, tight binding model 
\item[MSC:] 17B61, 17B68, 81R50, 81R60
\end{description}

%
%

%
\newpage
\setcounter{page}{1}
\setcounter{footnote}{0}
\setcounter{equation}{0}
\setcounter{secnumdepth}{4}
\renewcommand{\thepage}{\arabic{page}}
\renewcommand{\thefootnote}{\arabic{footnote})}
\renewcommand{\theequation}{\thesection.\arabic{equation}}

\section{Introduction}
\subsection{Background}
\label{sec:BG}
\indent
Noncommutative geometry is an interesting subject in the field of physics. 
It includes the theory of noncommutative fields that occur under certain antisymmetric fields~\cite{NCFT,SW}, as well as the description of noncommutative manifolds that correspond to quantum Hall states~\cite{QHE,GMP}. AdS/CFT provides a strong correspondence between theories related to black holes and quantum Hall effects~\cite{AdS}. This has led to interesting discussions such as the correspondence between edge states and black hole entropy~\cite{Edge}.

Moreover, there is an interesting topic in quantum gravity that involves infinite-dimensional symmetries. Moyal deformations of self-dual gravity have recently been studied in the context of noncommutativity and $W_{1+\infty}$ algebra~\cite{BHS}. On the other hand, the classical counterpart $w_{1+\infty}$ algebra~\cite{Winf} is related to soft graviton symmetries in asymptotically flat 4D quantum gravity~\cite{Strom}
\beq
[w_n^p,w_m^q]=(n(q-1)-m(p-1))w_{n+m}^{p+q-2}\,.  \label{winf1}
\eeq

While in the field of quantum Hall physics, 
another type of $W_{1+\infty}$ algebra has been studied in both the conformal field theory 
of edge excitations and the extension of bulk physics~\cite{latestQH}. 
It is also known that this algebra is related to the area-preserving 
diffeomorphisms of incompressible fluids~\cite{APD}, and its generators $\tilde{w}_n^k$ satisfy the commutation relation~\cite{CTZ,CTZ2}:
\beq
[\tilde{w}_n^k, \tilde{w}_m^l]=((k+1)(m+1)-(n+1)(l+1))\tilde{w}_{n+m}^{k+l}\,. \label{winf2}
\eeq
The generating functions $\rho(k,\bar{k})$ consisting of $\tilde{w}_n^k$ 
as Fourier components~\cite{CTZ2} obey the 
Girvin-MacDonald-Platzman algebra~\cite{GMP}, or rather
the Fairlie-Fletcher-Zachos (FFZ) type commutation relation~\cite{FFZ}:
\beq
[W_{k,\bar{k}},W_{p,\bar{p}}]=2\sinh(\frac{p\bar{k}-\bar{p}k}{8})W_{k+p, \bar{k}+\bar{p}}\,,
\eeq
by changing the normalization
\beq
\rho(k,\bar{k})=e^{-\frac{k\bar{k}}{8}}W_{k,\bar{k}}\,.
\eeq\par

The Moyal sine algebra, also known as the FFZ algebra, is associated with the Moyal bracket deformation. This deformation is a Lie-algebraic deformation of the Poisson brackets. The Moyal bracket and its star product are defined as follows~\cite{Moyal}: 
\begin{align}
& \{f(x,p),g(x,p)\}_\ast =\frac{2}{\hbar}\sin(\frac{\hbar}{2}\theta^{ab}\partial^a_1\partial^b_2)f(x,p)g(x,p)\,,\\
&f*g=\exp{i\frac{\hbar}{2}\theta^{ab} \partial_1^a \partial_2^b }f(x,p)g(x,p)\,,
\label{f*g}
\end{align}
where $\theta^{xp}=-\theta^{px}=-\theta$, and $\partial_1$ and $\partial_2$ denote 
forward (left) and backward (right) derivative operations, respectively. 
The Moyal commutator is related to the Moyal bracket
\beq
[f,g]_\ast=f\ast g - g\ast f =i\hbar \{f,g\}_\ast  \label{fg*bracket}
\eeq
and is often referred to as Moyal deformation or Moyal quantization. It can be 
regarded as a quantum deformation of the Dirac bracket $ i\hbar \{f,g\}_\ast \ra [f,g]$. 
This quantization leads to the $SU(\infty)$ Lie algebra, so-called 
the Moyal sine algebra~\cite{FFZ}
\beq
[\tau_{n,k}, \tau_{m,l}] = 2i \sine(\frac{\hbar\theta}{2}(nl-mk)) \tau_{n+m,k+l} \,, \label{Msine}
\eeq
if one takes the basis of $T^2$ phase space
\beq
\tau_{n,k}=e^{i(nx+kp)} \,. \label{txp}
\eeq\par

In the realm of mathematical notions of quantum groups, there is a foundational concept known as the covariance of noncommutative spaces, or quantum spaces~\cite{ma,tak,maj,wz,glq}. These have yielded significant results in solvable lattice models and YB equations related to quantum algebras~\cite{DJ,skr,frt,wor}. Quantum algebras are dual to quantum groups. 
Recently, attempts~\cite{JS,KS,AS2} to construct algebraic generators 
using noncommutative translation operations, which manifest the properties of 
noncommutative coordinates, have revealed the quantum space interpretation of 
the Moyal type deformed commutation relations, represented by the 
Hom-Lie algebras~\cite{Hom1,Hom2,Hom3}. These originate in the 
Curtright-Zachos (CZ) algebras~\cite{CZ,AS,NQ} 
(see also \cite{superCZ,superCZ2} and references in \cite{AS2}). 
This conceptual development has highlighted a new aspect of magnetic translation (MT)
operators within the context of tight binding models (TBM)~\cite{WZ,FK,HKW,MB}. 
Specifically, it emphasizes the newfound capability for quantum space interpretation, suggesting a correspondence between Wyle matrix representations~\cite{NQ} and 
$q$-difference MT operators. 

The $q$-difference MT operator is a noncommutative translation operator that satisfies the FFZ and the CZ algebra (Hom-Lie-Virasoro algebra) as a linear combination. In the discrete model TBM, the operational behavior of MT's matrix representation reveals that the phase shift arising from the path dependence of translational operations between two points on the plane leads to the quantum space noncommutativity of the TBM in a square lattice space~\cite{AS2}. The phase-shifted commutation relations of the CZ algebra can be explained by this path-dependent phase difference based on Wyle matrix representations. In other words, introducing phase-shifting products and changing the commutators to phase-shifted ones as quantum plane effects lead to the derivation of the Hom-Lie-Virasoro algebra. Moreover, the CZ generator is known to manifest in quantum space~\cite{superCZ,superCZ2}, rendering the projection of MT onto quantum space a compelling prospect. This feature arises from the matrix representation of MT exhibiting commutation relations analogous to the operational relations of quantum space.

\subsection{Motivation}
\label{sec:Aim}
\indent

In the previous paper \cite{AS2}, a $\ast$-bracket for the CZ algebra was 
defined using MT operators that have the properties of the Moyal bracket, 
but it remains an issue that the correspondence with the Moyal $\ast$-bracket 
is opaque. Namely, the algebraic relation of the CZ algebra was 
expressed borrowing the same notation as \eqref{fg*bracket},
\beq
[L_n,L_m]_\ast=(L_nL_m)_\ast-(L_mL_n)_\ast=[n-m]L_{n+m}\,,  \label{CZ} 
\eeq
where $(L_nL_m)_\ast=q^{m-n}L_nL_m$ and the $q$-bracket symbol $[A]$ is defined as
\beq
[A]=\frac{q^A -q^{-A}}{\qqi}\,,\quad  \mbox{where}\quad q=e^{i\hbar\theta} \,, 
\label{qbraA}
\eeq
however it is not clear whether $(L_nL_m)_\ast$ is equal to $L_n\ast L_m$
in the sense of \eqref{fg*bracket}. 
If we can extend the form that retains the Moyal product structure of the MT operator while inheriting its properties, we may be able to grasp the relationship 
between the $\ast$-bracket $(L_nL_m)_\ast$ and the Moyal product $L_n\ast L_m$. 
When such an extension is achieved, it will be possible to consider how the quantum 
space structure of the CZ generators is manifested, and how to understand 
the mysteries concerning associativity and nontrivial Hopf algebra problems. 

This time, we will explore the extension of the CZ algebra from two perspectives. 
The first perspective is to consider an extension that applies a scale operator to 
the CZ generators. This structure involves that the MT operators of noncommutative modes are constructed by multiplicative operations of the scale operator to those 
of trivial commutative modes. This scale operator is a central element of the CZ algebra~\cite{AS}.

The second perspective is to decompose the extended operator linearly and 
examine the role of each term. We will show that when viewed as a 
$W_\infty$-like multi-mode Hom-Lie deformation from a single CZ mode, 
it corresponds to a linear combination of \eqref{winf1} and \eqref{winf2} 
that appear in quantum gravity and quantum Hall effect, respectively.
We will also confirm that the same extension structure holds in the MT matrix representation. We then define the analytic form of the $\ast$-bracket product 
for each Hom-Lie type deformation, 
and show that these are derived by projecting from the Moyal product.

The organization of this paper is as follows. 
In Section~\ref{sec:CZ}, we provide a brief overview of the $CZ$ algebra and 
MT operators from the previous paper, and explain the necessary notation. 
In Section~\ref{sec:GCZ}, we define new generators $L_n^{\pm(k)}$ by applying 
the scaling operator to the $CZ$ generators and present the general algebraic 
relations with an external phase factor in Section~\ref{sec:GCZdef}.

In Section~\ref{sec:GCZ*alg}, we choose special external phase factors to 
construct new extended algebras $\mathcal{CZ}^\pm$ and $\mathcal{CZ}^\ast$, 
and define the $\ast$-bracket in the extended operator set $\mathscr{M}_L$. 
In the previous paper, the weight of the $\ast$-bracket in the operator 
set $\mathscr{M}_T$ of $CZ^\ast$ was fixed to 2, but in the extended algebra 
$\mathcal{CZ}^\ast$, the weight is extended to $k$. 
Because these algebras have different ways of taking weights, we explain how 
$CZ^\ast$ is embedded in $\mathcal{CZ}^\ast$ in Section~\ref{sec:subAlg}.

In Section~\ref{sec:winf}, changing our perspective on $\mathcal{CZ}^\ast$ 
from the scaling operator multiplication to the linear decomposition 
as mentioned above, we derive two Hom-Lie type deformations of the $w_{1+\infty}$ 
algebras. The first type, $W_\infty^\pm$, corresponds to the deformation 
of \eqref{winf1}, and is discussed in Section~\ref{sec:W1pm}. 
The second type, $\tilde{W}_\infty^\pm$, to the deformation of \eqref{winf2} 
is presented in Section~\ref{sec:W2pm}. For a consistent formulation of 
the $\ast$-bracket, we present the respective intersecting algebras 
$W_\infty^\ast$ and $\tilde{W}_\infty^\ast$ in Section~\ref{sec:W*}.

In Section~\ref{sec:XYrep}, we explore the connection between the extended algebra 
$\mathcal{CZ}^\ast$ and physical models. By using the correspondence between the matrix representation of MT and the $q$-difference representation, we confirm that the same extension structure is present in the matrix representation. Also, we obtain the matrix representations of two kinds of Hom-Lie generators, namely $\mathcal{CZ}^\ast$ and $W_\infty^\ast$. Following this, we consider a superposition of a Hermite operator based on an analogy of the hopping integral in solid-state physics~\cite{kittel}. We show an example of the TBM Hamiltonian that can be expressed in terms of $\mathcal{CZ}^-$ and $CZ^-$ generators.

In Section~\ref{sec:*CZW}, we discuss the projection relationship with the Moyal bracket. In Section~\ref{sec:*expr}, we obtain the analytical expression of the $\ast$-bracket product of each Hom-Lie type deformation, including $CZ^\ast$. Then, in Section~\ref{sec:*moyal}, we demonstrate that these $\ast$-brackets can be formulated as a projection from the Moyal $\ast$-product. In particular, $\mathcal{CZ}^\ast$ has a 1:1 equivalence relationship, while others require a reduction measure that enforces a weight setting in the Moyal operator \eqref{f*g}. Finally, we show that the second type $\tilde{W}_\infty^\ast$ is related to the first type via a coordinate transformation in their Moyal phase space.

%
\setcounter{equation}{0}
\section{CZ algebra}
\label{sec:CZ}
\indent

The CZ algebra \eqref{CZ} is a Hom-Lie deformation of the Virasoro algebra, and it 
is shown to be obtained by the FFZ generators~\cite{AS2}. There also discussed two 
other algebras $CZ^-$ and $CZ^\ast$ in addition to \eqref{CZ}, the $CZ^+$ algebra. 
The algebras $CZ^\pm$ are symmetric in the interchange of $q\lra q^{-1}$, and 
the $CZ^\ast$ algebra is an extended algebra composed of $CZ^\pm$. 
What we call the FFZ algebra here is originally given by \eqref{Msine} with the 
introduction of the deformation parameter $q$ and the $q$-bracket defined in \eqref{qbraA}. 
Changing the normalization
\beq
T_{(n,k)} =\frac{1}{\qqi} \tau_{n,k}  \label{T2base}
\eeq
we have the FFZ algebra
\beq
[T_{(n,k)},T_{(m,l)}]=[\frac{nl-mk}{2}] T_{(n+m,k+l)}\,, \label{FFZsine}
\eeq
where we assume that $\tau_{n,k}$ is generalized to an arbitrary operator behaving 
like the Moyal star products~\cite{FFZ} 
\beq
\tau_{n,k}\tau_{m,l} =q^{\frac{nl-mk}{2}} \tau_{n+m,k+l} \,. \label{FFZstar}
\eeq
It is well known that magnetic translation (MT) operators satisfy this 
fusion relation~\cite{Zaq}.

In this paper, we deal with the angular momentum representation instead of \eqref{txp} (see \S\ref{sec:*moyal}), where the MT operator $\tau_{n,k}$ is given by $\Op{t}{n}{k}$ 
as follows: 
\begin{eqnarray}
\Op{t}{n}{k} &=&
z^n q^{-k(z\partial +\frac{n}{2}+\Delta)} =e^{n\ln{z}}e^{-ia^2l_B^{-2}k(z\partial +\frac{n}{2}+\Delta)} \nn\\
&=&\Exp{\frac{i}{\hbar}\bm{\lambda}\cdot\bm{\Theta}}  \label{Tmtheta}
\end{eqnarray}
with the vectors $\bm{\lambda}$ and $\bm{\Theta}$
\beq
\bm{\lambda}=(nl_B,k\frac{a^2}{l_B})\,,
\eeq
\beq
(\Theta_1,\Theta_2)=(\frac{\hbar}{l_B}\theta, -\frac{1}{l_B}\mathcal{J}_3 
-\frac{\hbar}{l_B}\Delta)\,,\quad\, \mathcal{J}_3=-i\hbar\partial_\theta\,,\quad
z=e^{i\theta}\,, \label{varset}
\eeq
where we have introduced the magnetic length $l_B=\sqrt{\hbar c/eB}$ and 
a unit length $a$. The deformation parameter $q$ is given by the relation 
\beq
q=\exp{ia^2l_B^{-2}}\,.
\eeq
We should note that $\theta$ in \eqref{varset} is different from the one in \eqref{qbraA}. 
The phase space of $\Op{t}{n}{k}$ is represented by $\mathcal{J}_3$ and $\theta$, 
and the commutation relations
\beq
[\mathcal{J}_3,\theta]=-i\hbar\,,\quad [\Theta_i,\Theta_j]=-i\frac{\hbar^2}{l_B^2}\eps^{ij}\,.
\label{TTcom}
\eeq

Introducing the scaling operator $\hat{S}_0$ and the normalized MT operator 
$\hat{T}_n^{(k)}$ 
\beq
\hat{S}_0=q^{-2z\partial}\,,
\eeq
\beq
\Op{t}{m}{k} = z^m q^{-k(z\partial + \frac{m}{2} + \Delta)} =(\qqi)\Op{T}{m}{k}\,, \label{That} 
\eeq
we can rewrite $\Op{T}{n}{k}$ in the form based on $\Op{T}{n}{0}$ multiplied 
by $\hat{S}_0$ from left
\beq
\Op{T}{n}{k}=q^{k(\frac{n}{2}-\Delta)}\hat{S}_0^{\frac{k}{2}}\Op{T}{n}{0}\,.  \label{Tnk2}
\eeq

The $CZ^\pm$ generators are defined by the MT operators~\cite{AS2}
\beq
\hat{L}_n^\pm = \mp\Op{T}{n}{0} \pm q^{\pm(n+2\Delta)} \Op{T}{n}{\pm2}\,,  \label{Lpm1} 
\eeq
and satisfy the following $\ast$-bracket commutation relations:
\begin{align}
&    [\Op{T}{n}{k},\Op{T}{m}{l}]_\ast =0\,,                 \label{*1}     \\
& [\hat{L}_n^\pm, \hat{L}_m^\pm]_\ast = [n-m]\hat{L}_{n+m}^\pm\,,   \label{*2}   \\
&  [\hat{L}_n^\pm,\Op{T}{m}{l}]_\ast = -[m]\Op{T}{n+m}{l}\,.  \label{*3}  
\end{align}
The $CZ^\ast$ algebra, interacting algebras among $CZ^\eps$ generators 
($\eps=\pm$), reads 
\beq
 [L_n^\eps,L_m^\eta]_\ast =q^{\eta m}[n]L_{n+m}^\eps - q^{\eps n}[m]L_{n+m}^\eta\,,  \label{CZCZ}
\eeq
and the $\ast$-bracket commutator
\beq
[X_n^{(k)},X_m^{(l)}]_\ast=(X_n^{(k)}X_m^{(l)})_\ast-(X_m^{(l)}X_n^{(k)})_\ast \label{*4}
\eeq
is defined for every element $X_n^{\eps(k)} \in \mathscr{M}_T\{\hat{L}_n^\eps,T_n^{(k)}\} $ 
of weight $k$ by
\beq
(X_n^{\eps(k)}X_m^{\eta(l)})_\ast = q^{-x(\eps,\eta)} X_n^{\eps(k)} X_m^{\eta(l)}\,,
\quad\,\,  x(\eps,\eta)=\frac{\eta nl-\eps mk}{2}  \label{X*X}
\eeq
where $k$ and $l$ in $x(\eps,\eta)$ should be set to 2 for $\hat{L}_n^\eps$ and $k$ 
for $\hat{T}_n^{(k)}$. The signature symbol $\eps$ means $\pm$ for $\hat{L}_n^\pm$ 
and $\eps=+$ for $\hat{T}_n^{(k)}$.
In addition, $\hat{S}_0$ plays the role of central element if we regard 
the weight of $\hat{S}_0^k$ as $2k$ in \eqref{X*X} for $k\not=0$~\cite{AS} 
\beq
[\hat{S}_0^k, \hat{L}_n]_\ast =0\,,\quad [\hat{S}_0^k, \hat{T}_m^{(l)}]_\ast =0\,,
\eeq
and note that
\beq
\hat{S}_0=1+(q-q^{-1})\hat{L}_0\,.
\eeq
Since the $CZ^\pm$ generator \eqref{Lpm1} can be written in the $q$-difference 
form by using \eqref{That}
\beq
\hat{L}_n^\pm = \mp z^{n}\frac{1-q^{\mp2z\partial}} {\qqi} \,,  \label{Ln_diff}
\eeq
we refer to this as $q$-difference (MT) representation as well.

We have another representation, the cyclic matrix representation~\cite{NQ} 
with parameters $a_\pm$ and $b$, 
\beq
{\mathrm L}^\pm_n=\mp\left( \frac{1-{\mathrm Q}^{\pm2}}{q-q^{-1}}
+A_n^{\pm} {\mathrm Q}^{\pm2} \right)
{\mathrm H}^n \,,\quad A_n^{\pm}=a_\pm+b(q^{\pm2n}-1)\,,    \label{QHCZ}
\eeq
in terms of Wyle base matrices ${\mathrm H}$ and ${\mathrm Q}$ having the 
commutation relation
\beq
 {\mathrm H} {\mathrm Q}= q{\mathrm Q} {\mathrm H}\,,
\eeq
where
\beq
H_{jk}=\delta_{k+1, j}\,,\quad Q_{jk}=q^{j-1}\delta_{jk}\,,\quad\mbox{for}\quad j,k \in [1,N] \quad(\mbox{mod}\,N) \,.
\eeq
If we choose $a_\pm$ and $b$ to be
\beq
a_\pm=0\,,\quad b=-1/(\qqi)\,,
\eeq
and applying the correspondence to \eqref{QHCZ}
\beq
z \lra {\mathrm H}\,,\quad q^{\mp2z\partial} \lra {\mathrm Q}^{\pm2} \,,
\label{MT2HQ}
\eeq
we can verify that it coincides with the $q$-difference representation \eqref{Ln_diff}.

%
%
\setcounter{equation}{0}
\section{Scaled CZ algebras}
\label{sec:GCZ}
\indent

In this section, we discuss the extension of the CZ operators with the same structure 
as \eqref{Tnk2}. The structure is such that $\Op{T}{n}{0}$ is given a $k$-dependence 
through the scaling operator $\hat{S}_0$, resulting in $\Op{T}{n}{k}$. 
Applying this structure to the CZ generators $\Op{L}{n}{0}$ replacing $\Op{T}{n}{0}$ 
in \eqref{Tnk2},  
\beq
\Op{L}{n}{k}=q^{k(\frac{n}{2}-\Delta)}\hat{S}_0^{\frac{k}{2}}\Op{L}{n}{0}  \label{LSL}
\eeq
we consider the algebra of generalized CZ operators $\Op{L}{n}{k}$. 

Section~\ref{sec:GCZdef} presents a general description of the results to ensure broad 
applicability. The initial stage of the study should be conducted in a general framework 
to maximize flexibility and accommodate future adjustments. We therefore focus on 
the MT operator based on the abstract $T_n^{(k)}$ that only imposes the fusion rule 
and $q$-inversion symmetry, rather than the specific expression \eqref{That}.
We also introduce the following phase-shifted commutation relation such that 
we can later adjust the phase factor $x$, instead of the $\ast$-bracket 
commutator \eqref{*4} with no flexibility:
\beq
[X_n^{\eps(k)}, X_m^{\eta(l)}]_{(x)}=
q^xX_n^{\eps(k)}X_m^{\eta(l)}-q^{-x}X_m^{\eta(l)}X_n^{\eps(k)} \,.
\eeq

In Section~\ref{sec:GCZ*alg}, choosing the phase $x$ as
\beq
x=\frac{1}{2}(\eta nl-\eps mk)\,, \label{xphase}
\eeq
we present a generalized algebra $\mathcal{CZ}^\ast$ with $\ast$-product structure. 

In Section~\ref{sec:subAlg}, we derive the algebra $CZ^\ast$ as a subalgebra 
of $\mathcal{CZ}^\ast$ by setting the shifting weights $k$ and $l$ as 2 in \eqref{xphase}
\beq
x=\eta n-\eps m \,.
\eeq
Assigning $k=l=2$ is not a straightforward solution, and so we provide a comprehensive 
explanation.

\subsection{Definition}
\label{sec:GCZdef}
\indent

Let us define $L_n^{\pm(k)}$ of its shifting weight $k$ 
by generalizing $\hat{L}_n^\pm$ of weight $\pm2$ defined in \eqref{Lpm1}
\beq
L_n^{+(k)}=-\VT{n}{-k}+g_+ q^{n+2\Delta}\VT{n}{-k+2}\,,\quad\quad
 L_n^{-(k)}=\VT{n}{k}-g_- q^{-n-2\Delta}\VT{n}{k-2}\,,    \label{def_Ln+k}
\eeq
such that the $CZ^\pm$ modes $L_n^\pm$ are included as
\beq
\Lp{n}{0}{\pm} = L_n^\pm\,.
\eeq
We have introduced $g_\pm$ the parameters for later convenience, satisfying the 
relation
\beq
g_+g_-=1\,,
\eeq
and obeying the $q$-inversion symmetry rule
\beq
g_+\,\lra\,g_-\,.
\eeq

Suppose $T_n^{(k)}$ satisfy the following fusion rule
\beq
T_n^{(k)}T_m^{(l)}= \frac{1}{\qqi} q^{\frac{nl-mk}{2}}  T_{n+m}^{(k+l)}\,,  \label{Trans}
\eeq
and the $q$-inversion symmetry rules
\beq
\VT{n}{k} \lra -\VT{n}{-k}\,,\quad  L_n^{+(k)} \lra  L_n^{-(k)}\,.  \label{phaseinv}
\eeq
These rules can be verified by the real expression for MT operator \eqref{That}.

The upper index of $L_n^{\pm(k)}$ in \eqref{def_Ln+k} is by definition shifted 
by $k$ from $L_n^{\pm}$. 
As can be seen from the MT expression \eqref{Tnk2}, 
the $k$-shift is realized by the operator $S_0^{\frac{k}{2}}$ on $\VT{n}{0}$. 
If we write down $L_n^{\pm(k)}$ in terms of the MT differential operator \eqref{That}, 
we can verify the same relation as \eqref{LSL} (up to the sign factor on $k$):
\beq
\hat{L}_n^{\pm(k)}= q^{\mp k(\frac{n}{2}-\Delta)} \hat{S}_0^{\mp\frac{k}{2}} 
\hat{L}_n^\pm  \,,   \label{LnkbyLn}
\eeq
\beq
\hat{S}_0^\pm=1\pm(\qqi)\hat{L}_0^\pm=q^{\mp2z\partial}\,,
\eeq 
As can be seen from \eqref{LnkbyLn}, the $k$-dependence of $\hat{L}_n^{\pm(k)}$ is 
induced by the action of the scaling operator $\hat{S}_0$,  and 
$L_n^{-(k)}$ or $L_n^{+(-k)}$ show the same $k$-dependence 
as that of $\hat{T}_n^{(k)}$ in \eqref{Tnk2}. 
This common scaling structure between $L_n^{(k)}$ and $T_n^{(k)}$ may suggests  
the existence of a common phase factor in $\ast$-bracket 
without introducing the asymmetric weight setting such as \eqref{X*X},  
that is, $\pm2$ for $L_n^\pm$, and $k$ for $T_n^{(k)}$.

The phase-shifted commutation relations of $\Lp{n}{k}{\pm}$ for parameters 
$p$ and $r$ are derived from \eqref{Trans} as follows:
\begin{align}
&\qCom{\Lp{n}{k}{+}}{ \Lp{m}{l}{+}}{p}=[x-p]\Lp{n+m}{k+l}{+} 
+ g_+ q^{2\Delta}[n-m-x+p]\Lp{n+m}{k+l-2}{+}\,, \label{L++}\\
&\qCom{\Lp{n}{k}{+}}{\Lp{m}{l}{-}}{r}=q^n[n-y-r]\Lp{n+m}{l-k}{-} - q^{-m}[m-y-r]\Lp{n+m}{k-l}{+}\,,   \label{L+-}
\end{align}
where $x$ and $y$ are given by
\beq
x=\frac{nl-mk}{2}\,,\quad\quad y=\frac{nl+mk}{2}\,.
\eeq
The remaining commutators 
\begin{align}
&\qCom{\Lp{n}{k}{-}}{\Lp{m}{l}{-}}{-p}=[x-p]\Lp{n+m}{k+l}{-} 
+ g_- q^{-2\Delta}[n-m-x+p]\Lp{n+m}{k+l-2}{-}\,, \label{L--}\\
&\qCom{\Lp{n}{k}{-}}{\Lp{m}{l}{+}}{-r}=q^{-n}[n-y-r]\Lp{n+m}{l-k}{+}  - q^{m}[m-y-r]\Lp{n+m}{k-l}{-}\,, \label{L-+}
\end{align}
can be obtained from \eqref{L++} and \eqref{L+-} by applying the $q$-inversion 
symmetry formula
\beq
\pqCom{A(q)}{B(q)}{p}{r} \quad \lra\quad \pqCom{A(q^{-1})}{B(q^{-1})}{-p}{-r}\,.
\eeq

Here, we notice the following $k$-dependency structure
\beq
\Lp{n}{2-k}{\pm}=g_\pm q^{\pm(n+2\Delta)}\Lp{n}{k}{\mp}\,. 
\label{kinv}
\eeq
The formula \eqref{phaseinv} expresses the inversion of the signature $\pm$ on $L_n^{\pm(k)}$ under the exchange $q\lra q^{-1}$, while this formula also supplies an inversion relation concerning the sign of weight $k$ as well as the 
dependency relationship. 
Reducing the number of degrees of freedom using the $k$-dependency relationship \eqref{kinv} can be an effective approach depending on the nature and context of the problem. However, it is important to be aware of the potential drawbacks, 
such as increased algebraic complexity and reduced insight into symmetry. 
Thus, to preserve insights into symmetry, it is desirable to implement the reduction 
at the later stages of a computational process. 

At the present stage, without fixing the values of $p$ and $r$, we denote the generalized algebra of operators obtained by scaling from $CZ$ generators as $\mathcal{CZ}$ ( or $\mathcal{CZ}_{p,r}$ if necessary), into which the $k$-dependence is then introduced by the scaling operation. 
In the following subsections, setting specific values of $p$ and $r$, we obtain some characteristic algebras of $\Lp{n}{k}{+}$, $\Lp{n}{k}{-}$ and $\Lp{n}{k}{\pm}$, denoting 
these as $\mathcal{CZ}^+$, $\mathcal{CZ}^-$ and ${\mathcal{CZ}}^\ast$ respectively. 
We investigate the $\ast$-bracket structure and the $CZ^\ast$ subalgebra of the 
$\mathcal{CZ}$ algebras.

In Section \ref{sec:GCZ*alg}, setting $(p,r)=(x,-y)$, we derive the $\mathcal{CZ}^\pm$ and ${\mathcal{CZ}}^\ast$ that possess a $\ast$-bracket structure (denoting the generator set as $\mathscr{M}_L$ instead of $\mathscr{M}_T$ appeared in \eqref{X*X}). 

In Section \ref{sec:subAlg}, setting $(p,r)=(m-n, m+n)$ and applying the inversion formula \eqref{kinv}, we show that the minimal subalgebras are $CZ^+$ or $CZ^-$, 
and then obtain $CZ^\ast$ as a subalgebra. We discuss that 
the $\ast$-bracket structure of $\mathscr{M}_L$ is the extension 
that essentially contains the $\ast$-bracket \eqref{X*X} of $CZ^\ast$ algebra.

\subsection{${\mathcal{CZ}}^\ast$ algebra and its $\ast$-product}
\label{sec:GCZ*alg}
\indent

Consider the case $(p,r)=(x,-y)$ as a special situation. 
In this case, not only the first terms on the right-hand sides of \eqref{L++} and \eqref{L--} vanish, but also the parameter $\Delta$ can be eliminated by an appropriate transformation as shown later: 
\begin{align}
&\qCom{\Lp{n}{k}{+}}{ \Lp{m}{l}{+}}{x}= g_+ q^{2\Delta}[n-m]\Lp{n+m}{k+l-2}{+}\,, \label{L++2} \\
&\qCom{\Lp{n}{k}{-}}{\Lp{m}{l}{-}}{-x}= g_- q^{-2\Delta}[n-m]\Lp{n+m}{k+l-2}{-}\,, \label{L--2} \\
&\qCom{\Lp{n}{k}{+}}{\Lp{m}{l}{-}}{-y}=q^n[n]\Lp{n+m}{l-k}{-} - q^{-m}[m]\Lp{n+m}{k-l}{+}\,.   \label{L+-2} \\
&\qCom{\Lp{n}{k}{-}}{\Lp{m}{l}{+}}{y}=q^{-n}[n]\Lp{n+m}{l-k}{+}  - q^{m}[m]\Lp{n+m}{k-l}{-}\,. \label{L-+2}
\end{align}

In the first place, we are going to introduce a $\ast$-bracket similar to \eqref{X*X} 
for the generator set ${\mathscr M}_L=\{ \Lp{n}{k}{+}, \Lp{n}{k}{-}  \}$. 
In the case of ${\mathscr M}_T=\{X_n^{(k)}| L_n^\pm, T_n^{(k)}  \}$, 
the weights of $L_n^\pm$ are set to specific integers $k=\pm2$ to ensure 
consistency with $T_n^{(k)}$ in the $\ast$-bracket \eqref{X*X}. 
On the other hand, in the case of ${\mathscr M}_L$ in which does not include $T_n^{(k)}$, 
we naturally expect that the weight of $L_n^{\pm(k)}$ can be set to $k$ without exception for the $k=0$ $CZ$ mode $L_n^{\pm(0)}=L_n^\pm$. 

Let us define the $\ast$-bracket that is consistent with Eqs.\eqref{L++2}-\eqref{L-+2} 
one by one for the possible $\pm$ combinations: 
\begin{align}
&(\Lp{n}{k}{+}\Lp{m}{l}{+})_\ast = q^x \Lp{n}{k}{+}\Lp{m}{l}{+}\,,\quad
(\Lp{n}{k}{-}\Lp{m}{l}{-})_\ast = q^{-x} \Lp{n}{k}{-}\Lp{m}{l}{-}\, \\
&(\Lp{n}{k}{+}\Lp{m}{l}{-})_\ast = q^{-y} \Lp{n}{k}{+}\Lp{m}{l}{-}\,,\quad
(\Lp{n}{k}{-}\Lp{m}{l}{+})_\ast = q^y \Lp{n}{k}{-}\Lp{m}{l}{+}\,
\end{align}
or rather, introducing $\eps$ and $\eta$ to represent arbitrary $\pm$ signs, 
the $\ast$-bracket of $\mathscr{M}_L$ can be expressed in a single formula as follows: 
\beq
(L_n^{\eps(k)} L_m^{\eta(l)})_\ast =q^{x(\eps,\eta) } L_n^{\eps(k)} L_m^{\eta(l)} \,,\quad 
x(\eps,\eta)=\frac{\eta nl-\eps mk}{2}\,.     \label{ML*}
\eeq
Then we can express Eqs.\eqref{L++2}-\eqref{L-+2} as
\begin{align}
&[L_n^{\eps(k)}, L_m^{\eps(l)}]_\ast  = g_\eps q^{2\eps\Delta}[n-m]L_{n+m}^{\eps(k+l-2)}\,, \quad(\eps=\pm) \label{LLast}\\
&[L_n^{\eps(k)}, L_m^{\eta(l)}]_\ast  = q^{\eps n}[n]L_{n+m}^{\eta(l-k)} - q^{\eta m}[m]L_{n+m}^{\eps(k-l)}\,. 
\quad(\eps\not=\eta) \label{LLpmast}
\end{align}
It may seem contradictory that the weight of $L_n^\pm$ in the CZ algebra is $\pm2$, 
while the same operator $L_n^{\pm(0)}$ possesses the weight 0. 
However, it is important to note that the phase factor becomes 1 when $k=l=0$ in \eqref{ML*}, which means that the algebraic structure differs from that of the CZ algebra. 
Therefore, there is no contradiction. 

Now we introduce the following normalized operator $\tilde{L}_n^{\pm(k)}$ in order 
to remove the $\Delta$ dependence on the right-hand side of \eqref{LLast}
\begin{align}
\tilde{L}_n^{\eps(k)}&=g_{-\eps}q^{-2\eps\Delta}L_n^{\eps(k)} \\
&=-\eps g_{-\eps}q^{-2\eps\Delta}\VT{n}{-\eps k}+\eps q^{\eps n}\VT{n}{-\eps k+2\eps}\,. \label{newL}
\end{align}
Since this transformation evokes extra $\Delta$ dependence in \eqref{LLpmast}, 
we further apply the $k$-dependency formula \eqref{kinv}
\beq
\Lp{n}{2-k}{\eps}=g_\eps q^{(n+2\Delta)\eps}\Lp{n}{k}{\eta}\,,\quad \mbox{for }\,\eps\not=\eta
\eeq
or
\beq
q^{2\eps\Delta}\tilde{L}_n^{\eps(2-k)}=g_\eta q^{\eps n}\tilde{L}_n^{\eta(k)}\,,\quad \mbox{for }\,\eps\not=\eta
\eeq
to \eqref{LLpmast} in order to eliminate the additional $\Delta$ that arises from 
the transformaion \eqref{newL}. After all getting rid of $g_\eps$ from \eqref{LLast} 
as well, we have
\begin{align}
&[\tilde{L}_n^{\eps(k)}, \tilde{L}_m^{\eps(l)}]_\ast  = [n-m]\tilde{L}_{n+m}^{\eps(k+l-2)}\,, \quad(\eps=\pm) \label{tL++}\\
&[\tilde{L}_n^{\eps(k)}, \tilde{L}_m^{\eta(l)}]_\ast  = q^{\eta m}[n]\tilde{L}_{n+m}^{\eps(k-l+2)} 
- q^{\eps n}[m]\tilde{L}_{n+m}^{\eta(l-k+2)}\,, \quad(\eps\not=\eta)  \label{tL+-}
\end{align}
Decomposing the structure constant on the right-hand side of \eqref{tL++} into 
\beq
[n-m]=q^{\eps m}[n]-q^{\eps n}[m]\,,  \label{[n-m]}
\eeq
both \eqref{tL++} and \eqref{tL+-} are finally organized into the following single form
\beq
[\tilde{L}_n^{\eps(k)}, \tilde{L}_m^{\eta(l)}]_\ast  
= q^{\eta m}[n]\tilde{L}_{n+m}^{\eps(k+\eps\eta l-2\eps\eta)} 
- q^{\eps n}[m]\tilde{L}_{n+m}^{\eta(l+\eps\eta k-2\eps\eta)}\,, \label{tLtL}
\eeq
where the $\ast$-bracket is given by \eqref{ML*}. The 1st and 2nd terms on RHS 
of \eqref{tLtL} are symmetric in $n\lra m$, $k\lra l$ and $\eps\lra\eta$, and the 
structure constants are presented in very similar forms as in \eqref{CZCZ}. 
We express this unified algebra \eqref{tLtL} as $\mathcal{CZ}^\ast$, and 
each of \eqref{tL++} as $\mathcal{CZ}^\pm$. 

Setting $k=l=2$ in \eqref{tLtL} we can formally obtain the same algebra as 
\eqref{CZCZ}
\beq
[\tilde{L}_n^{\eps(2)}, \tilde{L}_m^{\eta(2)}]_\ast  
= q^{\eta m}[n]\tilde{L}_{n+m}^{\eps(2)} 
- q^{\eps n}[m]\tilde{L}_{n+m}^{\eta(2)}\,, \label{tLtLsub}
\eeq
however we should remember that the sign of phase factor $x(\eps,\eta)$ in 
the $\ast$-products of $\mathscr{M}_L$ and $\mathscr{M}_T$ are 
opposite. Up to this difference the $CZ^\ast$ algebra \eqref{CZCZ} is 
included in the generalized CZ algebra $\mathcal{CZ}^\ast$. 
This difference is shown to be inessential in the last part of the following section.

\subsection{Subalgebras $CZ^\pm$ and $CZ^\ast$}
\label{sec:subAlg}
\indent

In this section, we examine why \eqref{tLtLsub} has a reversed-phase 
compared to \eqref{CZCZ}. To this end, going back to the algebras 
\eqref{L++}, \eqref{L+-}, \eqref{L--} and \eqref{L-+}, let us consider the 
possible subalgebras of $\mathcal{CZ}$ generators with weights 0 and 2. 

Depending on the possible choice of $(k,l)$ combination for $k,l=0,2$, 
we consider specific subalgebras with setting $(p,r)$ values provided by 
\beq
\beta=m-n\,, \quad \gamma=\frac{n+m}{2}\,.  \label{betagamma}
\eeq
The first type is the case $(k,l)=(0,0)$. In this case, we have the following three 
closed subalgebras of $L_n^{\pm(0)}$ for the choice $(p,r)=(\beta,\gamma)$,
\beq
\qCom{L_n^+}{L_m^+}{\beta} =[n-m]L_{n+m}^+\,,\quad   
\qCom{L_n^-}{L_m^-}{-\beta} =[n-m]L_{n+m}^-\,, \label{minL++}
\eeq
\beq
\qCom{L_n^+}{L_m^-}{\gamma}=[\frac{n-m}{2}]\left(q^nL_{n+m}^- + q^{-m}L_{n+m}^+\right)\,.   \label{minL+-}
\eeq
These subalgebras are obviously the closed sets $\{L_n^+\}$, $\{L_n^-\}$, 
and $\{L_n^\pm \}$ respectively. 

The next type $(k,l)=(2,2)$ with the choice of $(p,r)=(\beta,-\gamma)$ leads to 
the following subalgebras of weight 2 generators $L_n^{\pm(2)}$,
\beq
\qCom{L_n^{+(2)}}{L_m^{+(2)}}{\beta} =g_+ q^{2\Delta}[n-m]L_{n+m}^{+(2)}\,,\quad   
\qCom{L_n^{-(2)}}{L_m^{-(2)}}{-\beta} =g_- q^{-2\Delta}[n-m]L_{n+m}^{-(2)}\,,\label{L2+sub}
\eeq
\beq
\qCom{L_n^{+(2)}}{L_m^{-(2)}}{-\gamma}=[\frac{n-m}{2}]\left(q^nL_{n+m}^- + q^{-m}L_{n+m}^+\right)\,. \label{LLsub22}
\eeq
There are clearly the two sets $\{L_n^{+(2)}\}$ and $\{L_n^{-(2)}\}$ for \eqref{L2+sub}, 
and we see that \eqref{LLsub22} leads to the closed set $\{\Lp{n}{2}{\pm},L_n^\pm \}$ 
as seen below. 

The third combination is $(k,l)=(2,0)$. In this case, we have the following 
closed subalgebras of $L_n^{\pm(2)}$ and $L_n^{\pm(0)}$, with the choice 
$p=-\gamma$ in \eqref{L++} and \eqref{L--}:
\begin{align}
&\qCom{L_n^{+(2)}}{L_m^{+}}{-\gamma}=[\frac{n-m}{2}] \left( \Lp{n+m}{2}{+} 
+ g_+ q^{2\Delta}L_{n+m}^+ \right)\,, \label{LLsub20-}\\
&\qCom{L_n^{-(2)}}{L_m^{-}}{\gamma}=[\frac{n-m}{2}] \left( \Lp{n+m}{2}{-} 
+ g_- q^{-2\Delta}L_{n+m}^- \right)\,, \label{LLsub20+}
\end{align}
which generate the sets $\{\Lp{n}{2}{+}, L_n^+\}$ and $\{\Lp{n}{2}{-}, L_n^-\}$ respectively.

There remains another type of algebras to be considered with different sign combinations 
$L_n^{\pm(2)}$ and $L_n^{\mp(0)}$. For the case where $(k,l)=(2,0)$ with $r=-\beta$ in \eqref{L+-}, we have:
\beq
\qCom{L_n^{+(2)}}{L_m^{-}}{-\beta}=q^{-m}[n-m]\Lp{n+m}{2}{+}\,,\quad \label{L20sub}
\eeq
and for the case $(k,l)=(0,2)$ with $r=\beta$ in \eqref{L+-}: 
\beq
\qCom{L_n^{+}}{L_m^{-(2)}}{\beta}=q^{n}[n-m]\Lp{n+m}{2}{-}\,. \label{L02sub}
\eeq
These relations along with \eqref{LLsub22} comprise the closed set $\{\Lp{n}{2}{\pm},L_n^\pm \}$. 

{}~From the above results, we have derived eight possible sets of subalgebras. 
However, taking the inversion formula \eqref{kinv} into account, we find that 
five of these sets are overcounted.  
In fact, $\Lp{n}{2}{\pm}$ and $\Lp{n}{0}{\pm}$ are related by the following relation:
\beq
\Lp{n}{2}{\pm} = g_\pm q^{\pm(n+2\Delta)}\Lp{n}{0}{\mp}\,.
\eeq
This means that \eqref{L2+sub}, \eqref{L20sub} and \eqref{L02sub} reduce to the $CZ^\pm$ algebras \eqref{minL++}, while \eqref{LLsub22}, \eqref{LLsub20-} and 
\eqref{LLsub20+} reduce to \eqref{minL+-} exactly. 

After all the first three algebras \eqref{minL++} and \eqref{minL+-} are the only 
independent subalgebras
\beq
\{L_n^+\}\,, \quad    \{L_n^-\}\,, \quad   \{L_n^\pm \}\,,  \label{3subsets}
\eeq
where the first two sets correspond to the $CZ^+$ and $CZ^-$ algebras. 
The algebra of the third set $\{L_n^\pm \}$ has different structure constants from the $CZ^\ast$ algebra as compared from \eqref{minL+-} and \eqref{CZCZ}. 
The reason is originated in the specific choice $r=\gamma$ 
that is simply convenient to factor out $[\frac{n-m}{2}]$ from the structure 
constants of intersecting algebras between $L_n^{+(k)}$ and $L_n^{-(k)}$ for $k=0,2$ 
as seen from \eqref{minL+-} to \eqref{LLsub20+}.

However, even in the case of $r=2\gamma$, the independent subsets \eqref{3subsets} 
can be verified in the same way as above, up to the differences of the structure constants. 
In this case, \eqref{minL+-} is replaced to the following subalgebra
\beq
\qCom{L_n^+}{L_m^-}{2\gamma}=q^{-m}[n]L_{n+m}^+ - q^{n}[m]L_{n+m}^-\,, \label{minL+-2}
\eeq
and using the $\ast$-bracket \eqref{X*X}, the algebras \eqref{minL++} and \eqref{minL+-2} are expressed as
\begin{align}
& [L_n^\eps,L_m^\eps]_\ast = [n-m] L_{n+m}^\eps\,, \\
& [L_n^\eps,L_m^\eta]_\ast =q^{\eta m}[n]L_{n+m}^\eps - q^{\eps n}[m]L_{n+m}^\eta\,. \quad (\eps\not=\eta)
\end{align}
Applying the formula \eqref{[n-m]}, these can further be organized into the $CZ^\ast$ 
form \eqref{CZCZ}, of which structure constants are the same as those of 
$\mathcal{CZ}^\ast$ algebra \eqref{tLtL}. 

As to \eqref{tLtLsub} the subalgebra of $\mathcal{CZ}^\ast$, noticing the following relations 
with the choice $g_\pm=1$ and $\Delta=0$, 
\beq
\tilde{L}_n^{\eps(k)}=L_n^{\eps(k)}\,,\quad
L_n^{\eps(2)}=q^{\eps n}L_n^{\eta(0)}\,,(\eps\not=\eta)
\eeq
we can recognize that $CZ^\ast$ is included in the subalgebras of $\mathcal{CZ}^\ast$, 
since $\tilde{L}_n^{\pm(2)}$ satisfies the same algebra of $L_n^{\mp(0)}=L_n^\mp$, that is 
$CZ^\mp$, keeping in mind the difference between the $\ast$-brackets 
\eqref{ML*} for $\mathscr{M}_L$ and \eqref{X*X} for $\mathscr{M}_T$. 
The difference means that the signs of $x(\eps,\eta)$ are opposite each other in 
$\mathcal{CZ}^\ast$ algebra \eqref{tLtL} and $CZ^\ast$ algebra \eqref{CZCZ}. 

This issue can be resolved by redefining the $\Lp{n}{k}{\pm}$ operator in 
the definition \eqref{def_Ln+k} by replacing the $k$-dependence with $-k$. 
In that case, \eqref{ML*} and \eqref{tLtL} are changed to the following:
\beq
(L_n^{\eps(k)} L_m^{\eta(l)})_\ast =q^{-x(\eps,\eta) } L_n^{\eps(k)} L_m^{\eta(l)} \,,\quad 
x(\eps,\eta)=\frac{\eta nl-\eps mk}{2}\,,     \label{ML*2}
\eeq
\beq
[\tilde{L}_n^{\eps(k)}, \tilde{L}_m^{\eta(l)}]_\ast  
= q^{\eta m}[n]\tilde{L}_{n+m}^{\eps(k+\eps\eta l-2\eps\eta)} 
- q^{\eps n}[m]\tilde{L}_{n+m}^{\eta(l+\eps\eta k-2\eps\eta)}\,. \label{tLtL2}
\eeq
While the $\ast$-bracket in both $\mathscr{M}_T$ and $\mathscr{M}_L$ now shares the same form given by equation \eqref{ML*2}, it is crucial to note that the treatment of weights for $CZ^\ast$ generators still differs depending on whether or not 
$\VT{n}{k}$ is included in the generator set. 
Specifically, the weight in $\mathscr{M}_T$ is fixed to $k=2$ for all $CZ^\ast$ generators, 
while in $\mathscr{M}_L$, the weight is simply $k$. 
Thus, setting $k=l=2$ in \eqref{tLtL2} reproduces \eqref{tLtLsub}, which is the same 
as the $CZ^\ast$ algebra \eqref{CZCZ}. 
Although setting $k=l=0$ in \eqref{tLtL} or \eqref{tLtL2} does not lead to \eqref{CZCZ}, 
we should stress that this does not mean a contradiction at all. 
It just means that $L_n^\eps$ is not only the element of $\mathcal{CZ}^\ast$ algebra 
\eqref{tLtL}, but also that of $CZ^\ast$ algebra \eqref{CZCZ}.

%
\setcounter{equation}{0}
\section{Hom-Lie type deformation of $w_{1+\infty}$}
\label{sec:winf}
\indent

In Section~\ref{sec:GCZ}, we introduced the ${\mathcal{CZ}}^\pm$ algebra 
from the perspective of deriving an algebra analogous to $CZ$. 
In this Section, we consider a different extension from the perspective of the 
Hom-Lie type deformation of $w_{1+\infty}$ mentioned in Section~\ref{sec:BG}. 
In the relevance of ${\mathcal{CZ}}^\pm$ algebras, 
the role of $\Delta$ in the definition \eqref{def_Ln+k} of $L_n^{\pm(k)}$ 
may be of some interest. 

To begin with, we are interested in the roles played by each term in the linear combination 
in the definition of $L_n^{\pm(k)}$ given by \eqref{def_Ln+k}. 
We then rewrite the definition as follows:
\begin{align}
&L_n^{\pm(k)}=\tilde{W}_n^{\pm(k)} - g_\pm q^{\pm(n+2\Delta)}W_n^{\pm(k)} \,, \\
&\tilde{W}_n^{\pm(k)}=\mp\VT{n}{\mp k}\,,\quad
 W_n^{\pm(k)} =\mp\VT{n}{\pm(2-k)}\,,
\end{align}
In our previous work~\cite{AS2}, we showed that for $k=0$, the pair 
$(\tilde{W}_n^{\pm(0)},W_n^{\pm(0)})$ consisted of a commutative $CZ$ representation 
$\VT{n}{0}$ and a non-commutative but $\ast$-bracket commutative $\VT{n}{\pm2}$. 
These two components combined to form the (non-commutative) $CZ^\pm$ operators. 
Our interest here lies in understanding the roles played by 
$\tilde{W}_n^{\pm(k)}$ and $W_n^{\pm(k)}$ when extended to $k\not=0$.

While the answer is not unique and hinges on the chosen $\ast$-bracket phases, this Section presents a possible approach where both $\tilde{W}n^{\pm(k)}$ and $W_n^{\pm(k)}$ emerge as independent deformations of $w{1+\infty}$, each with its own $\ast$-bracket. Significantly, they share the commutative $CZ$ representation as a shared subalgebra. 
Discussions on the $\ast$-product structure are deferred to Section~\ref{sec:*CZW}, while we first investigate the connection to TBM in Section~\ref{sec:XYrep}.

\subsection{ 1st type deformation ${W}^\pm_\infty$}
\label{sec:W1pm}
\indent

We now return to the general $\mathcal{CZ}^+$ algebra \eqref{L++} and 
consider a different value for the external bracket parameter $p$ 
than in Section~\ref{sec:GCZ}. First, we choose the $p$ value as
\beq
 p=\frac{\beta}{2}=\frac{m-n}{2}\,,   \label{choice1}
\eeq
and \eqref{L++} reads
\beq
\qCom{\Lp{n}{k}{+}}{ \Lp{m}{l}{+}}{\frac{m-n}{2}}  = [\frac{n(l+1)-m(k+1) }{2}]\Lp{n+m}{k+l}{+} 
- g_+ q^{2\Delta}[\frac{n(l-1)-m(k-1)}{2}]\Lp{n+m}{k+l-2}{+}\,. \label{L2++}\\
\eeq
Let us consider the situation that the second term on RHS of \eqref{L2++} becomes 
dominant, for example, the case $|q|>1$ with $\Delta\ra\infty$ (or $g_+\ra\infty$). 
To see the effect of this situation, using the decomposition
\beq
L_n^{+(k)}=-\VT{n}{-k} - g_+ q^{n+2\Delta}W_n^k \,,\quad  W_n^k =-\VT{n}{-k+2}\,,
\label{W+}
\eeq
and extracting the terms of $O(q^{4\Delta})$ from both sides of \eqref{L2++}, we obtain
\beq
\qCom{W_n^k}{W_m^l}{\frac{m-n}{2}} = [\frac{n(l-1)-m(k-1) }{2}] W_{n+m}^{k+l-2}\,.
\eeq

Similarly in the case of $\mathcal{CZ}^-$ algebra \eqref{L--} for $L_n^{-(k)}$, 
introducing the following notation with rewriting $W_n^k$ in \eqref{W+} as $W_n^{+(k)}$ 
\beq
L_n^{\pm(k)}=\mp\VT{n}{\mp k} - g_\pm q^{\pm(n+2\Delta)}W_n^{\pm(k)} \,,
\quad  W_n^{\pm(k)} =\mp\VT{n}{\mp(k-2)}\,, \label{Wpm}
\eeq
and extracting $O(q^{-4\Delta})$ in the limit $\Delta\ra-\infty$ with $|q|>1$, 
we have the algebra (denoting $W_\infty^\pm$)
\beq
\qCom{W_n^{\pm(k)}}{W_m^{\pm(l)}}{\pm\frac{m-n}{2}} 
= [\frac{n(l-1)-m(k-1) }{2}] W_{n+m}^{\pm(k+l-2)}\,. \label{WpmCom}
\eeq
These algebras $W_\infty^\pm$ are automorphic under the following 
transformations for an arbitrary parameter $\delta$:
\beq
\Wp{n}{k}{\pm'}=q^{\mp(k-2)\delta} \Wp{n}{k}{\pm}\,. \label{Wpm2}
\eeq

In order to see the correspondence between $W_\infty^\pm$ and \eqref{winf1} 
more clearly, let us redefine the deformation parameter $q$ as 
$q\ra q_1=q^{\frac{1}{2}}$, and introduce redefined bracket symbols
\beq
[A,B]'_{(\alpha)}=q_1^{\alpha}AB-q_1^{-\alpha}BA   \,,\quad
  [x]'=\frac{q_1^x -q_1^{-x}}{q_1-q_1^{-1}}\,.     \label{norm_w1}
\eeq
Changing the normalization from $W_n^{\pm(k)}$ to $w_n^{\pm(k)}$ such as 
\beq
   w_n^{\pm(k)}=[2]' W_n^{\pm(k)}\,,  \label{norm_w2}
\eeq
we obtain the following form similar to \eqref{winf1}
\beq
[ w_n^{\pm(k)}, w_m^{\pm(l)}]'_{\pm(m-n)} = [n(l-1)-m(k-1)]' w_{n+m}^{\pm(k+l-2)} \,.
\label{WpmCom2}
\eeq
This relation is obvious to reduce \eqref{winf1} as $q\ra1$. If we set $k=l=2$, 
there exists the special algebra of commutable $CZ^\pm$ representation $w_n^{\pm(2)}$, 
which plays the same role as $T_n^{(0)}$ in $CZ$ generators:
\beq
[ w_n^{\pm(2)}, w_m^{\pm(2)}]'_{\pm(m-n)} = [n-m]' w_{n+m}^{\pm(2)} \,.\label{CZsub1}
\eeq

At this stage, whether \eqref{WpmCom} or \eqref{WpmCom2} can be cast into the $\ast$-bracket formalism such as \eqref{CZ} and \eqref{*4} remains an open question. However, we will revisit this issue and discuss it in detail in Section~\ref{sec:*CZW}.

\subsection{2nd type deformation $\tilde{W}^\pm_\infty$}
\label{sec:W2pm}
\indent

Next, let us consider the case
\beq
 p =\frac{(m-l)-(n-k)}{2} \,.
\eeq
This case corresponds to the replacement $\beta\ra\beta+k-l$ in \eqref{choice1}, 
giving rise to the additional term $\frac{k-l}{2}$ on RHS of \eqref{choice1}. 
Substituting the above $p$ into \eqref{L++}, we have
\begin{align}
\qCom{\Lp{n}{k}{\pm}}{ \Lp{m}{l}{\pm}}{\pm p}  
&= [\frac{(n+1)(l+1)-(m+1)(k+1) }{2}]\Lp{n+m}{k+l}{\pm}  \nn\\
&- g_\pm q^{\pm2\Delta}[\frac{(n+1)(l-1)-(m+1)(k-1)}{2}]\Lp{n+m}{k+l-2}{\pm}\,. 
\label{L3++}
\end{align}
Let us consider the situation that the first term on RHS of \eqref{L3++} becomes 
dominant, for example, 
the case $|q|>1$ with $\Delta\ra\mp\infty$ (or $g_\pm\ra0$). 
Using the decomposition 
\beq
  L_n^{\pm(k)}= \tilde{W}_n^{\pm(k)} \pm g_\pm q^{\pm(n+2\Delta)}\VT{n}{\mp(k-2)} \,,\quad  \tilde{W}_n^{\pm(k)} =\mp\VT{n}{\mp k}\,, \label{tildeW}
\eeq
and extracting $O(1)$ terms on both sides of \eqref{L3++}, we obtain the following algebra (denoting $\tilde{W}_\infty^\pm$) as a deformation of \eqref{winf2},
\beq
\qCom{\tilde{W}_n^{\pm(k)}}{\tilde{W}_m^{\pm(l)}}{\pm p} =
 [\frac{(n+1)(l+1)-(m+1)(k+1) }{2}] \tilde{W}_{n+m}^{\pm(k+l)}\,, \label{tildeWCom}
\eeq
which has the following automorphism
\beq
\tilde{W}_{n}^{\pm'(k)}=q^{\mp k\delta} \tilde{W}_{n}^{\pm(k)}\,, \label{tildeW2}
\eeq
where $\delta$ is independent of the $W_\infty^\pm$ automorphism \eqref{Wpm2}. 

In order to clarify the correspondence between $\tilde{W}_\infty^\pm$ and \eqref{winf2}, 
we modify the notation following the same convention as in \eqref{norm_w1} 
and \eqref{norm_w2}, obtaining the following:
\beq
[ \tilde{w}_n^{\pm(k)}, \tilde{w}_m^{\pm(l)}]'_{(\pm2p)} =
 [(n+1)(l+1)-(m+1)(k+1)]' \tilde{w}_{n+m}^{\pm(k+l)} \,. \label{tildeWCom2}
\eeq
This relation is obvious to reduce \eqref{winf2} as $q\ra1$. If we set $k=l=0$, 
we can verify the commutable representation $\tilde{w}_n^{\pm(0)}$ satisfying 
the $CZ^\pm$ algebra
\beq
[ \tilde{w}_n^{\pm(0)}, \tilde{w}_m^{\pm(0)}]'_{\pm(m-n)} = [n-m]' \tilde{w}_{n+m}^{\pm(0)} \,.
\label{CZsub2}
\eeq

Here, focusing on $\mathcal{CZ}^+$, it is intriguing to observe that the deformation $W_\infty^+$ of \eqref{winf1} emerges at $\Delta\ra\infty$ (or $g_+\ra\infty$), 
describing high-energy physics, while the deformation $\tilde{W}_\infty^+$ 
of \eqref{winf2} appears at $\Delta\ra-\infty$ (or $g_+\ra0$), 
describing low-energy physics. This suggests that $\mathcal{CZ}^+$ can be viewed 
as a unified framework for handling operators at different scales. 
The situation is reversed for $\mathcal{CZ}^-$, where $\tilde{W}_\infty^-$ 
for low energy emerges at $\Delta\ra\infty$ (or $g_-\ra0$) and $W_\infty^-$ 
for high energy appears at $\Delta\ra-\infty$ (or $g_-\ra\infty$). 
Hence, it is reasonable to say that $W_\infty^\pm$ constitutes 
the high-energy algebra set and $\tilde{W}_\infty^\pm$
the low-energy algebra set.

\subsection{Intersecting algebras $W^\ast_\infty$ and $\tilde{W}^\ast_\infty$}
\label{sec:W*}
\indent

The algebra of interaction between $W_\infty^+$ and $W_\infty^-$ 
can be calculated by using \eqref{Wpm}
\beq
[W_n^{\eps(k)},W_m^{-\eps(l)}]_{(\frac{\eps}{2}(m+n))}=
[\frac{n(l-1)+m(k-1)}{2}]\frac{\eps}{2}
(W_{n+m}^{+(\eps k -\eps l +2)} -W_{n+m}^{-(-\eps k +\eps l +2)} )\,.
\eeq
Together with \eqref{WpmCom}, we can organize them into a unified expression
(denoting $W_\infty^{\ast}$)
\beq
[W_n^{\eps(k)},W_m^{\eta(l)}]_{(\frac{\eps m -\eta n}{2})}=
[\frac{n(l-1)-\eps\eta m(k-1)}{2}] \sum_{\nu=\pm}
 C^\nu_{\eps,\eta} W_{n+m}^{\nu(\eps\nu k +\eta\nu l -2\eps\eta)} \,,\label{Wpm*}
\eeq
where
\beq
C^\nu_{\eps,\eta}=\frac{1}{2}(1+\nu\eps)\delta_{\eps,\eta}
+\frac{1}{2}\nu\eps\delta_{\eps,-\eta}\,. \label{Cnu}
\eeq
If we set $k=l=2$, we obtain the following subalgebras:
\beq
[W_n^{\eps(2)},W_m^{\eps(2)}]_{(\eps\frac{m-n}{2})}
=[\frac{m-n}{2}]W_{n+m}^{\eps(2)}\,, \label{Wsub++}
\eeq
\beq
[W_n^{+(2)},W_m^{-(2)}]_{(\frac{n+m}{2})}=
\frac{1}{2}[\frac{n+m}{2}](W_{n+m}^{+(2)}-W_{n+m}^{-(2)})\,. \label{Wsub+-}
\eeq
While the first equation corresponds to the $CZ^\pm$ algebra as in \eqref{minL++} (as mentioned earlier), the second equation differs in its structure constants 
from \eqref{minL+-} and therefore does not correspond to $CZ^\ast$. 

In the same way as above, the intersecting algebra of $\tilde{W}_\infty^\pm$ 
can be calculated in terms of \eqref{tildeW},
\beq
[\tilde{W}_n^{\eps(k)}, \tilde{W}_m^{-\eps(l)}]_{(\frac{(\eps(m+l)+\eps(n+k)}{2})}=
[\frac{(n+1)(l+1)+(m+1)(k+1)-2}{2}]\frac{\eps}{2}
(\tilde{W}_{n+m}^{+(\eps k -\eps l )} -\tilde{W}_{n+m}^{-(-\eps k +\eps l )} )\,.
\eeq
Together with \eqref{tildeWCom}, we can encapsulate them into a unified form
(denoting $\tilde{W}_\infty^{\ast}$)
\beq
[\tilde{W}_n^{\eps(k)}, \tilde{W}_m^{\eta(l)}]_{(\tilde{p})}=
[\frac{(n+1)(l+1)-\eps\eta(m+1)(k+1)+\eps\eta-1}{2}]\sum_{\nu=\pm} 
C^\nu_{\eps,\eta} \tilde{W}_{n+m}^{\nu(\eps\nu k +\eta\nu l )} \,, \label{tildeW*}
\eeq
where $C^\nu_{\eps,\eta}$ is given by \eqref{Cnu}, and 
\beq
\tilde{p}=\frac{1}{2} \{\,  (\eps m -\eta l)-(\eta n-\eps k)\, \}\,.
\eeq
We can obtain the subalgebras if we set $k=l=0$ in \eqref{tildeW*}, however 
noticing the relation $\tilde{W}_n^{\eps(k)}=W_n^{\eps(k+2)}$, it is clear that 
one can derive \eqref{Wsub++} and \eqref{Wsub+-}, the same subalgebras of 
$W_\infty^\ast$.

Here, we remark on the non-triviality of $W_\infty^\ast$. 
While it may be argued that $W_\infty^\ast$ and $\tilde{W}_\infty^\ast$ 
are merely algebras of $T_n^{(k)}$ from \eqref{Wpm} and \eqref{tildeW}, 
the phase factors of the phase-shifted commutators of these algebras 
(left-hand sides of \eqref{Wpm*} and \eqref{tildeW*}) differ from those of the 
$\mathscr{M}_T$ algebra (see \eqref{*1}-\eqref{*3}). 
I particular, if we consider $\{W_n^{\eps(k)},\tilde{W}_n^{\eps(k)}\}:=Y_n^{(k)}$ 
as an element of $\mathscr{M}_T$, 
we can obtain equivalent relations to \eqref{*1} and \eqref{*3} 
by assigning the weight $\eps(2-k)$ to $W_n^{\eps(k)}$
and the weight $-\eps k$ to $\tilde{W}_n^{\eps(k)}$: 
\beq
[W_n^{\eps(k)},T_m^l]_{(\eps m-\frac{nl+\eps mk}{2})}=0\,,\quad
[L_n^\eps,W_m^{\eta(l)}]_{(\eps m-\eta n+\frac{\eta}{2}nl)}=-[m]W_{n+m}^{\eta(l)}\,,
\eeq
\beq
[\tilde{W}_n^{\eps(k)},T_m^l]_{(-\frac{\eta}{2}n-\frac{\eps}{2}mk)}=0\,,\quad
[L_n^\eps,\tilde{W}_m^{\eta(l)}]_{(\eps m+\frac{\eta}{2}nl)}=-[m]\tilde{W}_{n+m}^{\eta(l)}\,,
\eeq
and these can be written in terms of \eqref{X*X}, the $\ast$-product of $\mathscr{M}_T$:
\beq
[Y_n^{(k)},T_m^l]_\ast=0\,,\quad
[L_n^\eps,Y_m^{(l)}]_\ast=-[m]Y_{n+m}^{(l)}\,.
\eeq
However, \eqref{Wpm*} and \eqref{tildeW*} can not be written in terms of \eqref{X*X}. 
That means that the phase structure encoded in \eqref{Wpm*} and \eqref{tildeW*} 
can not be captured by the $\mathscr{M}_T$ framework. 

Therefore, $W_\infty^\ast$ and $\tilde{W}_\infty^\ast$ are distinct from the 
$\mathscr{M}_T$ algebra. For later convenience, we introduce the notations
$\mathscr{M}_W$ and $\mathscr{M}_{\tilde{W}}$ to represent the operator sets 
of these algebras. In $\mathscr{M}_W$ and $\mathscr{M}_{\tilde{W}}$, the weight of 
$Y_n^{(k)}$ will differ from the aforementioned values. 
(We should recall that the same operator can have different weight values depending on 
the algebra, as seen with $L_n^{\pm}$, which has weight $\pm2$ in $\mathscr{M}_T$ 
and weight 0 in $\mathscr{M}_L$.)

Up to this point, the choice of the external phase factors in the phase-shifted 
commutators has been motivated by a couple of subjective considerations, such as 
obtaining a natural form for $W_\infty^\pm$ 
and facilitating a unified expression of $W_\infty^\ast$. 
The justification for this choice as an objective $\ast$-product structure 
will be addressed in Section~\ref{sec:*CZW}.

%
\setcounter{equation}{0}
\section{Wyle representations and TBM Hamiltonian}
\label{sec:XYrep}
\indent

This section uses the cyclic matrix representations to explore the relationship between TBM and $W_\infty^\ast$ or $\tilde{W}_\infty^\ast$ algebras. Since $W_n^{\pm(k)}$, $\tilde{W}_n^{\pm(k)}$ and $L_n^{\pm(k)}$ are composed of $T_n^{(k)}$, their matrix representations are obtained from the differential operator $\Op{T}{n}{k}$ 
given by \eqref{That} or \eqref{Tnk2} through the replacement by \eqref{MT2HQ}, 
which is the correspondence between the Wyle matrices and the differential operators.
\beq
\Op{T}{n}{k}=\eqref{Tnk2} \quad\stackrel{\eqref{MT2HQ}}{\longrightarrow} \quad
{\mathrm T}_n^{(k)}=q^{-k(\frac{n}{2}+\Delta)} \frac{{\mathrm H}^n}{\qqi}
(q{\mathrm Q})^k\,. \label{matTnk}
\eeq
Without any specific reason, there is no necessity for the value of $\Delta$ 
to be identical in both representations. We hence distinguish them by denoting $\Delta$ 
as $\delta$ in the matrix representation, and we have chosen the value of $\delta$ 
in \eqref{matTnk} as 
\beq
\Delta\,\quad\ra\, \delta=\Delta-1\,.
\eeq

Substituting \eqref{matTnk} into \eqref{def_Ln+k}, we can find the matrix representation 
of $L_n^{\pm(k)}$,
\beq
{\mathrm L}_n^{\pm(k)}=\mp q^{\pm k(\frac{n}{2}+\Delta)}\frac{{\mathrm H}^n}{\qqi}
\left( (q{\mathrm Q})^{\mp k} - g_\pm (q{\mathrm Q})^{\mp(k-2)} \right)\,.
\label{Lnk_g}
\eeq
It is noteworthy that setting k=0 in this equation yields the $CZ^\pm$ generators. 
However, it is crucial to note that the correspondence with the differential operator  
representation requires the choice of $g_\pm=q^{\mp2}$ (which cancels the effect of 
shifting the $\delta$ value by 1). This choice corresponds to the setting 
$a_\pm=0$ and $b=-1/(\qqi)$ in \eqref{QHCZ}. 
Using the scaling operator of this case
\beq
{\mathrm S}_0^\pm={\mathrm Q}^{\pm2}=1\pm(\qqi){\mathrm L}_0^{\pm}\,,
\eeq
and
\beq
{\mathrm L}_n^\pm=\mp \frac{{\mathrm H}^n}{\qqi}
({\mathrm 1} - g_\pm q^{\pm2}{\mathrm Q}^{\pm2} ) \,,
\eeq
The $\mathcal{CZ}^\pm$ generators \eqref{Lnk_g} can be expressed in the form where the 
the scaling operator acts on the $CZ^\pm$ generators,
\beq
{\mathrm L}_n^{\pm(k)}= q^{\mp k(\frac{n}{2}-\Delta)} {\mathrm S}_0^{\mp\frac{k}{2}} 
{\mathrm L}_n^\pm  \,. \label{LkS0L}
\eeq
This relation also holds for the differential MT operator representation \eqref{That} 
(with $g_\pm=q^{\mp2}$). Writing $\hat{L}_n^\pm$ in terms of MT operator 
from \eqref{def_Ln+k} and \eqref{That}, we verify
\beq
\hat{L}_n^{\pm(k)}= q^{\mp k(\frac{n}{2}-\Delta)} \hat{S}_0^{\mp\frac{k}{2}} 
\hat{L}_n^\pm  \,,
\eeq
\beq
\hat{S}_0^\pm=1\pm(\qqi)\hat{L}_0^\pm=q^{\mp2z\partial}\,,
\eeq
where $\hat{L}_n^\pm$ is given by \eqref{Lpm1}. The key takeaway here is that 
the $k$-dependence of $CZ^\pm$ stems from the action of the scale operators. 
That is precisely why $CZ^\pm$ is considered a scaled CZ algebra.

As to $W^\ast_\infty$ and $\tilde{W}^\ast_\infty$, applying \eqref{matTnk} 
to \eqref{Wpm} and \eqref{tildeW}, we have
\beq
W_{n}^{\pm(k)}=\mp q^{\pm(k-2)(\Delta+\frac{n}{2})} \frac{{\mathrm H}^n}{\qqi}
(q{\mathrm Q})^{\mp (k-2)} \,, \label{WHQ*1}
\eeq
\beq
\tilde{W}_{n}^{\pm(k)}=\mp q^{\pm k(\Delta+\frac{n}{2})} \frac{{\mathrm H}^n}{\qqi}
(q{\mathrm Q})^{\mp k} \,. \label{WHQ*2}
\eeq
The difference of $\tilde{W}_n^{\pm(k)}$ from $W_n^{\pm(k)}$ is just the value of $k$, 
however, we are dealing with the different phase-shifted commutators 
\eqref{WpmCom2} and \eqref{tildeWCom2} as well as the interacting $\pm$ algebras 
\eqref{Wpm*} and \eqref{tildeW*}. Another remark is that 
setting $\nu=\Delta$ in \eqref{Wpm2} and \eqref{tildeW2} leads to $\Delta=0$ 
in above \eqref{WHQ*1} and \eqref{WHQ*2}, 
satisfying \eqref{Wpm*} and \eqref{tildeW*} of course.

\vskip\baselineskip
To construct a physical system described by ${\mathrm T}_n^{(k)}$,
we begin by reviewing the basic concept of TBM described by the matrices ${\mathrm H}$ and ${\mathrm Q}$. The model represents a two-dimensional lattice system of free electrons in a magnetic field, where particle hopping is expressed through creation and annihilation operators, and the AB phase is incorporated into the MT operator (hereafter referred to as the discrete MT representation). 
The Hamiltonian for TBM takes the following form,
where the discrete MT operator acts on a two-dimensional lattice:
\beq
H=\hat{T}_x+\hat{T}_y+\hat{T}^\dagger_x+\hat{T}^\dagger_y\,, \label{TBH}
\eeq
with
\beq
\hat{T}_x=\sum_{n,m}e^{i\theta_{mn}^x}c^\dagger_{m+1,n}c_{m,n}\,,\quad\quad 
\hat{T}_y=\sum_{n,m}e^{i\theta_{mn}^y}c^\dagger_{m,n+1}c_{m,n}\,,    \label{TBTxTy}
\eeq
where $\theta^x_{mn}$ and $\theta^y_{mn}$ are the $AB$ phases associated with 
a unit displacement in the $x$ and $y$ directions
\beq
\theta^x_{mn}: (m,n)\ra (m+1,n)\,,\quad \theta^y_{mn}: (m,n)\ra(m,n+1)\,.
\eeq
The matrix representation of the Hamiltonian is given by
\beq
H= i(X^{-1}-X)Y + iY^{-1}(X^{-1}-X)\,, \label{TBH2}
\eeq
with
\beq
X={\mathrm H}^{-1}\,,\quad Y=q{\mathrm Q}\,,
\eeq
leading to the following Schr\"{o}dinger equation
\beq
i(q^{j+1}+q^{-j})\psi_{j+1} - i(q^{-j}+q^{j-1)})\psi_{j-1} =E\psi_j\,.\label{MBeq}
\eeq
Based on the relation between \eqref{TBH2} and \eqref{MBeq}, we can generalize 
the system as follows: To derive the Schr\"{o}dinger equation for a system with an effective spacing of $n$ and a $k$-fold increase in the fluctuation phase of the quantum plane, we replace $\Delta j=1\ra n$ and $q\ra q^k$ in \eqref{MBeq}, 
\beq
i(q^{k(j+n)}+q^{-kj})\psi_{j+n} - i(q^{-kj}+q^{k(j-n)})\psi_{j-n} =E\psi_j \,.
\eeq
The generalized Hamiltonian, describing this equation, is then given by 
\beq
H_{(n,k)}= i(X^{-n}-X^{n})Y^k + iY^{-k}(X^{-n}-X^{n})\,. \label{Hnk}
\eeq
Using the matrix representation \eqref{matTnk}, we can rewrite $H_{(n,k)}$ as
\beq
H_{(n,k)}=i(\qqi)q^{k(\Delta+1)}(q^{\frac{nk}{2}}{\mathrm T}_n^{(k)} 
-q^{-\frac{nk}{2}}{\mathrm T}_{-n}^{(k)}) + \quad\mbox{h. c.}\,, \label{HbymatT}
\eeq
\beq
({\mathrm T}_{n}^{(k)})^\dagger=-{\mathrm T}_{-n}^{(-k)}\,.
\eeq
According to the relation $W_n^{\pm(k)}=\mp\VT{n}{\mp(k-2)}$, $H_{(n,k)}$ can be 
expressed by $W_\infty^\ast$ generator
\beq
H_{(n,k)}=i(\qqi)q^{k(\Delta+1)}(q^{\frac{nk}{2}} W_n^{-(k+2)} 
-q^{-\frac{nk}{2}} W_{-n}^{-(k+2)}) + \quad\mbox{h. c.}\,, \label{HbyW}
\eeq
\beq
(W_{n}^{+(k)})^\dagger=W_{-n}^{-(k)}\,.
\eeq
It is tempting to choose $\nu=\Delta+1$ in \eqref{tildeW2} to get rid of 
$q^{k\Delta}$ with the use of \eqref{tildeW}, and we also have another expression 
in terms of $\tilde{W}_\infty^\ast$ generator
\beq
H_{(n,k)}=i(\qqi)(q^{\frac{nk}{2}}\tilde{W}_{n}^{-'(k)}
-q^{-\frac{nk}{2}}\tilde{W}_{-n}^{-'(k)}) + \quad\mbox{h. c.}\,, \label{HbyW'}
\eeq
\beq
(\tilde{W}_{n}^{+'(k)})^\dagger=\tilde{W}_{-n}^{-'(k)}\,.
\eeq
However it is equivalent to redefine ${\mathrm T}_n^k$ absorbing the factor $q^{k\Delta}$ as 
\beq
{\mathrm T}_x=iq^{k\Delta}q^{\frac{nk}{2}}(\qqi){\mathrm T}_n^{(k)}\,,\quad
{\mathrm T}_y=-iq^{k\Delta}q^{-\frac{nk}{2}}(\qqi){\mathrm T}_{-n}^{(k)}\,. \label{Tredef}
\eeq
In this case, the Hamiltonian \eqref{HbymatT} becomes 
\beq
H_{(n,k)}={\mathrm T}_x+{\mathrm T}_y + \quad\mbox{h. c.}\,, \label{TBHnk}
\eeq
which is similar to the familiar form of TBM Hamiltonian \eqref{TBH}. 
The Hamiltonian takes the form of a trace around a rectangular path of size $(n,k)$ 
spanning in the horizontal and vertical directions. 
The orientations of ${\mathrm T}_x$ and ${\mathrm T}_y$ should be noted. 
The $x$ direction corresponds not to the $n$ axis, but rather to the $k-n$ axis, 
while the $y$ direction aligns with the $k+n$ axis, not the $k$ axis. 
(See also below \eqref{tau_decomp})

On this occasion, let us define $\tau_x$ and $\tau_y$ according to \eqref{Tredef}, 
replacing the matrix ${\mathrm T}_n^{(k)}$ with the differential operator 
$\hat{T}_n^{(k)}$, in order to discuss the correspondence of \eqref{TBHnk} 
to its differential operator representation. 
\beq
\tau_x=iq^{\frac{nk}{2}}\tilde{t}_n^{(k)}\,,\quad
\tau_y=-iq^{-\frac{nk}{2}}\tilde{t}_{-n}^{(k)}\,,
\eeq
where
\beq
\tilde{t}_n^{(k)}:=q^{k\Delta} \hat{t}_n^{(k)}=q^{k\Delta}(\qqi)\hat{T}_n^{(k)}\,,
\eeq
and $\hat{t}_n^{(k)}$ is given by \eqref{That}. The redefined MT operators $\tilde{t}_n^{(k)}$ 
satisfy the same operational rules as the original ones $\hat{t}_n^{(k)}$. 
Namely, the fusion rule and the Hermite conjugation (h.c.) relation are given by
\beq
\tilde{t}_n^{(k)}\tilde{t}_m^{(l)}=q^{\frac{nl-mk}{2}}\tilde{t}_{n+m}^{(k+l)}\,,\quad
(\tilde{t}_{n}^{(k)})^\dagger=\tilde{t}_{-n}^{(-k)}\,. \label{mrule1}
\eeq
However, in the differential representation, the relation of h.c. can only be confirmed 
on the unit circle $|z|=1$. Replacing the matrix ${\mathrm T}_n^{(k)}$ in \eqref{HbymatT} with the differential operator $\hat{T}_n^{(k)}$ yields the differential operator corresponding to \eqref{TBHnk}
\beq
\hat{H}_{(n,k)}=\tau_x+\tau_y+(\tau_x)^\dagger+(\tau_y)^\dagger\,. \label{tbhnk}
\eeq
Since $\tau_x$ and $\tau_y$ can be decomposed by the fusion rules as follows:
\beq
\tau_x=i\tilde{t}_n^{(0)}\tilde{t}_0^{(k)}\,,\quad
\tau_y=-i\tilde{t}_{-n}^{(0)}\tilde{t}_0^{(k)}\, \label{tau_decomp}
\eeq
we recognize that they correspond to the translation in each direction of 
$X_r\,(r=n-k)$ and $-Y_j\,(j=k+n)$ (see Figure 1 and 2 in~\cite{AS2} for details). 
When applying the quantum plane interpretation, the coordinate transformation
\beq
(n,k) \quad\ra\quad (r,j)=(n-k,k+n)\,, \label{qplane_tr}
\eeq
is suggested. This transformation is necessary for understanding the $\ast$-bracket 
product for $\tilde{W}_\infty^\ast$, to be examined later.

To find the Hamiltonian described by ${\mathrm L}_n^{\pm(k)}$, 
we apply a transformation by linear combination based on the system \eqref{TBHnk} represented by ${\mathrm T}_n^{(k)}$. 
The Hamiltonian \eqref{TBHnk} or \eqref{tbhnk} exhibits a structure represented as a trace around a rectangular path with a length and width of $(n,k)$. 
Hence, we consider a trace that goes around a farther ($k<0$) or closer ($k>0$) 
rectangle to the nucleus (central integral) by $\pm2$ in $k$. 
This structure provides an analogical representation of summing the central integral and the hopping integral in the crystal field Hamiltonian. 
In this way, we introduce the following linear 
combination as a candidate for the new Hamiltonian:
\beq
H_g = H_{(n,k)} + g H_{(n,k-2)}\,. \label{defHg}
\eeq
Here, $g$ is a real number because of the Hermiticity of $H_g$. $H_g$ can then be written as
\beq
H_g=i(\qqi)q^{k\Delta}(q^{\frac{nk}{2}}V_n^{(k)}-q^{-\frac{nk}{2}}V_{-n}^{(k)} )
+ \quad\mbox{h.c.}\, \label{HgbyV}
\eeq
where
\beq
V_n^{(k)}=\tilde{\mathrm W}_n^{-(k)} +g q^{-(n+2\Delta)}{\mathrm W}_n^{-(k)}\,.\label{Vnk}
\eeq
Setting $g=-1$ makes $V_n^{(k)}$ coincide with the $\mathcal{CZ}^-$ generator 
${\mathrm L}_n^{-(k)}$ with $g_\pm=1$ in \eqref{Lnk_g}. 
\beq
H_g=i(\qqi)q^{k\Delta}(q^{\frac{nk}{2}}\mathrm{L}_n^{(k)}-
q^{-\frac{nk}{2}}\mathrm{L}_{-n}^{(k)} )
+ \quad\mbox{h.c.}\,,  \label{HgbyL-}
\eeq
where Hermitian conjugation of ${\mathrm L}_n^{\pm(k)}$: 
\beq
({\mathrm L}_n^{\pm(k)})^\dagger=q^{\mp kn}\tilde{\mathrm L}_{-n}^{\mp(k)}\,, \label{Lnkhc}
\eeq
with the definition
\beq
\tilde{\mathrm L}_{n}^{\pm(k)}={\mathrm H}^{-n}{\mathrm L}_{n}^{\pm(k)} {\mathrm H}^n\,.
\eeq
{}~Furthermore putting $k=0$, it reduces $CZ^-$ generator, obtaining the Hamiltonian
\beq
H_g=i(\qqi)({\mathrm L}_n^- - {\mathrm L}_{-n}^-)+ \quad\mbox{h. c.}\,\label{Hg}
\eeq
where
\beq
{\mathrm L}_n^\pm=\mp X^{-n} \frac{1-Y^{\pm2}}{\qqi}\,,\label{HgL}
\eeq
\beq
({\mathrm L}_n^\pm)^\dagger=\tilde{{\mathrm L}}_{-n}^\mp\,,\quad\quad
\tilde{{\mathrm L}}_{n}^\pm=X^n{\mathrm L}_{n}^\pm X^{-n}\,.
\eeq
Here is a comment on the relevance of the previous paper. 
This $CZ^-$ matrix representation \eqref{HgL} coincides with the equation (4.24) 
presented in~\cite{AS2}. However, $H_g$ does not coincide with $H_{(n,2)}$, which is 
discussed as a TBM Hamiltonian family in the previous paper (see equation (5.52)).

As another candidate for the Hamiltonian, let us consider replacing the central integral 
term with a different Hermitian operator
\beq
H'_g=H'_{(n,k)} + g H_{(n,k-2)}\,, \label{Hg'}
\eeq
where
\beq
H'_{(n,k)}=(\qqi) q^{k\Delta}(q^{\frac{nk}{2}} \tilde{\mathrm W}_n^{-(k)}- 
q^{-\frac{nk}{2}} \tilde{\mathrm W}_{-n}^{-(k)}  ) + \quad\mbox{h.c.}\,. \label{Hnk'}
\eeq
Introducing an operator similar to \eqref{Vnk}, we rewrite $H'_g$ as
\beq
H'_g=(\qqi)q^{k\Delta}(q^{\frac{nk}{2}}v_n^{(k)}-q^{-\frac{nk}{2}}v_{-n}^{(k)} )
+ \quad\mbox{h.c.}\,, \label{Hgbyv}
\eeq
where
\begin{align}
v_n^{(k)}&=\tilde{\mathrm W}_n^{-(k)} +ig q^{-(n+2\Delta)}{\mathrm W}_n^{-(k)}\,,
\label{vnk} \\
(v_n^{(k)})^\dagger&=\tilde{\mathrm W}_{-n}^{+(k)} 
-ig^\ast q^{n+2\Delta}{\mathrm W}_{-n}^{+(k)}\,. \label{vnkd}
\end{align}
Although $g$ is a real number, i.e., $g=g^\ast$, we have formally introduced 
the notation $g^\ast$ for later convenience. 
The replacement $g\ra ig$ yields the correspondence of $V_n^{(k)}$ to $v_n^{(k)}$. 
Using the matrix representation \eqref{WHQ*1}, \eqref{WHQ*2} and \eqref{Lnk_g}, 
we derive the Hamiltonian $H_g'$ described by the $\mathcal{CZ}^-$ generators
\begin{align}
H'_g&= {\mathrm H}^n(q{\mathrm Q})^k\{1+ig(q{\mathrm Q})^{-2}\}
+ (q{\mathrm Q})^{-k} \{1-ig^\ast(q{\mathrm Q})^{2}\} {\mathrm H}^{-n} 
- (n\ra -n)\\
&=(\qqi)(q^{k(\frac{n}{2}+\Delta)} {\mathrm L}_n^{-(k)} 
-q^{k(-\frac{n}{2}+\Delta)} {\mathrm L}_{-n}^{-(k)}) +\,\,\mbox{h.c.}\,, \label{Hg2byL}
\end{align}
where we have chosen $g_\pm$ as
\beq
g_+=ig^\ast\,,\quad g_-=-ig\,. \label{gpm_ig}
\eeq
Note that we have assumed in\eqref{Lnkhc}
\beq
g_\pm^\ast=g_\mp\,, \label{gpm*}
\eeq
and this is consistent with the choice \eqref{gpm_ig}. In the case of $k=0$, 
the Hamiltonian \eqref{Hg2byL} reduces to the form given by $CZ^-$ generators
\beq
H'_g=(\qqi)({\mathrm L}_n^- -{\mathrm L}_{-n}^-) +  \quad\mbox{h.c.}\,.
\label{Hg'byL}
\eeq
Remember that the $CZ^-$ generators in this equation are more general than 
in \eqref{Hg} because we have not fixed $g_\pm$ yet. We should note that 
$iH'_g=H_g$ does not hold by comparing the h.c. terms between $H_g$ \eqref{Hg} and 
$H'_g$ \eqref{Hg'byL} because they differ in their sign each other. However, 
$H'_g$ \eqref{Hg'byL} coincides with $H_g$ \eqref{Hg} if we apply the replacement 
$\mathrm{L}_n^\pm\ra \mp i\mathrm{L}_n^\pm$ to $H'_g$.

In the case of $k\not=0$, $H'_g$ \eqref{Hgbyv} transforms into $H_g$ \eqref{HgbyV} 
according to 
\beq
v_n^{(k)}\, \ra \, iV_n^{(k)}\,, \quad 
(v_n^{(k)})^\dagger\, \ra \, -i(V_n^{(k)})^\dagger\,.  \label{transV}
\eeq
In particular, in the case of $\mathrm{L}_n^{\pm(k)}$, since the generators of 
$H_g$ \eqref{HgbyL-} are chosen as $g=-1$ and $g_\pm=1$, we have to choose $g=i$ 
to match $H'_g$ \eqref{Hg2byL} to the same generators ($g_\pm=1$), 
so that the transformation \eqref{transV} reduces to 
\beq
\mathrm{L}_n^{-(k)}\, \ra\, i \mathrm{L}_n^{-(k)}\,,\quad
\tilde{\mathrm{L}}_n^{+(k)}\, \ra\,  -i \tilde{\mathrm{L}}_n^{+(k)}\,,
\label{transL}
\eeq
leading to $H'_g\ra H_g$. Note that taking $g=i$ violates Hermiticity requirements, 
but its significance will be illustrated in the following example. 
Depending on the appropriate choice of $g_\pm$, $H'_g$ can be expressed by 
various representations of ${\mathrm L}_n^{\pm(k)}$. 
Here are three examples for different values of $g$, 
including the case $g=i$ as well as $g=\pm1$. 
All these cases satisfy the condition \eqref{gpm*}. 

The first case is the imaginary value $g=i$, although it is forbidden by Hermiticity of 
$H'_g$ as mentioned above. Nevertheless, this case interestingly reproduces the 
first Hamiltonian $H_g$ in \eqref{Hg}, giving rise to the overall $i$ factor, 
which resembles the Wick rotation, though $iH'_g\not\propto H_g$. 
The operator \eqref{vnk} reads
\beq
v_n^{(k)}=\tilde{\mathrm W}_n^{-(k)} - q^{-(n+2\Delta)}{\mathrm W}_n^{-(k)}\,,
\label{vnk1}
\eeq
which reproduces the $\mathcal{CZ}^-$ matrix representation \eqref{Lnk_g}
with $g_\pm=1$:
\beq
v_n^{(k)}={\mathrm L}_n^{-(k)} \quad\,(\mbox{for}\quad g_\pm=1)\,.\label{vki}
\eeq
Setting $k=0$ reproduces the $CZ^-$ generator \eqref{HgL} 
participating in the Hamiltonian \eqref{Hg}, 
which is however different from $H_{(n,2)}$ by definition. 
%
Explicitly writing down the h.c. parts of \eqref{Hg} and \eqref{Hg'byL}, 
\beq
H_g=i(\qqi)\{(\mathrm{L}_n^{-}-\mathrm{L}_{-n}^{-})
-(\tilde{\mathrm{L}}_n^{+}-\tilde{\mathrm{L}}_{-n}^{+})\}\,,  \label{Hghc}
\eeq
\beq
H'_g=(\qqi)\{(\mathrm{L}_n^{-}-\mathrm{L}_{-n}^{-})
+(\tilde{\mathrm{L}}_n^{+}-\tilde{\mathrm{L}}_{-n}^{+})\}  \,, \label{Hghc'}
\eeq
it is clear that $iH'_g = H_g$ does not hold, since we can see that the relative signs 
of the $\mathrm{L}_n^-$ and $\tilde{\mathrm{L}}_n^+$ parts in these equations 
are opposite each other. Substituting \eqref{HgL}, we have
\begin{align}
&H_g=H_{(n,0)}-H_{(n,-2)}=i(\mathrm{H}^{n}-\mathrm{H}^{-n})(1-Y^{-2})
+i(1-Y^2)(\mathrm{H}^{n}-\mathrm{H}^{-n})\,, \\
&H'_g=iH_{(n,-2)}=(\mathrm{H}^{-n}-\mathrm{H}^{n})Y^{-2}
+Y^2(\mathrm{H}^{n}-\mathrm{H}^{-n})\,,
\end{align}
thus indicating $iH'_g \neq H_g$ with their difference 
$2i(\mathrm{H}^n - \mathrm{H}^{-n})$. 

As can be seen by comparing \eqref{Hghc} and \eqref{Hghc'}, 
we need the h.c. parts $H_g$ and $H'_g$ to have the same sign for $iH'_g = H_g$ to hold. 
So, if we apply a transformation that reverses this sign, $H'_g$ will be 
mapped into $H_g$. In fact, in the expression \eqref{Hghc'} for $H'_g$, 
if we redefine $\mathrm{L}_n^\pm$ as
\beq
\mathrm{L}_n^\pm \quad\ra\quad \mathbb{L}_n^\pm=\mp i\mathrm{L}_n^\pm\,,
\label{L2L} 
\eeq
then the h.c. part of $H'_g$ changes its sign with reproducing $H_g$. Hence, we find
\beq
H'_g(\mathbb{L}_n^\pm) = H_g(\mathrm{L}_n^\pm)\,. \label{H2H}
\eeq
The coefficient of $i\mathrm{L}_n^+$ in \eqref{L2L} is $-1$ because $\tilde{\mathrm{L}}_n^+$ transforms as the h.c. counterpart of $\mathrm{L}_n^-$, 
which is not an independent transformation arising from $\mathrm{L}_n^-$. 
Thus, the essential operation is simply multiplying $\mathrm{L}_n^{-(k)}$ by $i$. 

Although the Hamiltonian does not have an overall $i$ factor, the relation \eqref{H2H} implies that$\mathrm{L}_n^-$ as the fundamental constituent unit is related to an $i$ factor. 
The transformation \eqref{L2L} allows us to interpret the unphysical $H'_g$ with $g=i$ 
as the Hamiltonian of an imaginary world obtained by phase-rotation of 
the real Hamiltonian $H_g$ with $g=-1$. 
In the case of $k\neq 0$, the situations for \eqref{transV} and \eqref{transL} 
are the same. 

The second example is the case $g=1$. The operator \eqref{vnk} reads
\beq
v_n^{(k)}=\tilde{\mathrm W}_n^{-(k)} +i q^{-(n+2\Delta)}{\mathrm W}_n^{-(k)} \,,
\label{vnk2}
\eeq
which reproduces the $\mathcal{CZ}^-$ matrix representation ${\mathscr L}_n^{-(k)}$ 
given by substituting $g_\pm=\pm i$ in \eqref{Lnk_g}
\beq
v_n^{(k)}={\mathscr L}_n^{-(k)}\,,\quad\mbox{where}\,\,\,
{\mathscr L}_n^{\pm(k)}:={\mathrm L}_n^{\pm(k)}\quad
\mbox{for}\,\, g_\pm=\pm i\,. \label{vk+}
\eeq
Setting $k=0$ in $v_n^{(k)}$ reproduces the $CZ^-$ generator ${\mathscr L}_n^-$ 
(Eq.(4.21) in \cite{AS2}), which satisfies
\beq
({\mathscr L}_n^\pm)^\dagger=\tilde{\mathscr L}_{-n}^\mp\,,\quad
\tilde{\mathscr L}_{n}^\pm=X^n{\mathscr L}_{n}^\pm X^{-n}
=\pm\frac{1\mp iY^{\pm2}}{\qqi}X^{-n}\,, \label{CZtL}
\eeq
and $H'_g$ becomes $H_{(n,-2)}$,
\beq
H'_g=(\qqi)({\mathscr L}_n^- -{\mathscr L}_{-n}^-) +  \quad\mbox{h.c.} =H_{(n,-2)} \,.
\eeq
In addition, $H_{(n,2)}$ is given by ${\mathscr H}_n$ (Eq.(5.55) in \cite{AS2} and Appendix), 
which is obtained by replacing 
${\mathscr L}_n^-$ with the $CZ^+$ operator ${\mathscr L}_n^+$ in $H'_g$
\beq
{\mathscr H}_n=(\qqi)({\mathscr L}_n^+ -{\mathscr L}_{-n}^+) +  \quad\mbox{h. c.} 
=H_{(n,2)} \,.
\eeq

The third case is $g=-1$, and the operator \eqref{vnk} reads
\beq
v_n^{(k)}=\tilde{\mathrm W}_n^{-(k)} -i q^{-(n+2\Delta)}{\mathrm W}_n^{-(k)}\,,
\label{vnk3}
\eeq
which reproduces the $\mathcal{CZ}^-$ matrix representation ${\mathscr K}_n^{-(k)}$ 
given by substituting $g_\pm=\mp i$ in \eqref{Lnk_g}
\beq
v_n^{(k)}={\mathscr K}_n^{-(k)}\,,\quad
{\mathscr K}_n^{\pm(k)}:={\mathrm L}_n^{\pm(k)}\quad\mbox{for}\quad g_\pm=\mp i\,,
\label{vk-}
\eeq
Setting $k=0$, we can verify that $H'_g$ describes $H_{(-n,-2)}$ in terms of the 
$CZ^-$ generator ${\mathscr K}_n^-:={\mathscr K}_n^{-(0)}$,
\beq
H'_g=(\qqi)({\mathscr K}_n^- -{\mathscr K}_{-n}^-) +  \quad\mbox{h. c.}=H_{(-n,-2)}\,,
\eeq
where
\beq
{\mathscr K}_n^\pm=\mp X^{-n}\frac{1\pm i Y^{\pm2}}{\qqi}\,,
\label{CZKn}
\eeq
\beq
({\mathscr K}_n^\pm)^\dagger=\tilde{\mathscr K}_{-n}^\mp\,,\quad
\tilde{\mathscr K}_{n}^\pm=X^n {\mathscr K}_{n}^\pm X^{-n}\,.
\eeq
Note that ${\mathscr K}_n^\pm$ satisfy the $CZ^\ast$ algebra.

%
\setcounter{equation}{0}
\section{The $\ast$-product formulation}
\label{sec:*CZW}
\indent

In the previous paper, we interpreted the $\ast$-bracket in CZ algebra as 
an ordered product on the quantum plane, and the phase generated by 
the $\ast$-bracket as a phase difference caused by the fluctuation 
of the path between two points connected by the translation operator 
on the quantum plane.

Despite the similarity between the $\ast$-bracket and the Moyal bracket 
\eqref{fg*bracket}, we have considered them to be different because of the unclear 
relationship between the $\ast$-bracket and its classical analytical definition like 
the Moyal $\ast$-product \eqref{f*g}. 
In this Section, we will discuss the relationship between them. 
In the subsequent discussion, the $\ast$-bracket may be referred to as the $\ast$-ordered product or simply $\ast$-product, particularly when emphasizing its connection to the Moyal $\ast$-product.

In Section~\ref{sec:*expr}, we derive an analytical expression for the 
$\ast$-ordered product as preparation for exploring the relationship 
with the Moyal $\ast$ factor. This expression is similar to the definition of the 
$\ast$-product in the Moyal bracket. 

Section~\ref{sec:*moyal} derives our $\ast$-bracket factor from the Moyal 
bracket factor. In the case of $\mathcal{CZ}^\ast$, the $\ast$-bracket exactly 
coincides with the Moyal $\ast$-product, while others, $CZ^\ast$ and the 
two types of $W_\infty^\ast$, need a reduction measure imposing weight settings 
in the Moyal operator. We show that the second type $\tilde{W}_\infty^\ast$ is 
related to the first type via a specific re-combination of Moyal phase space. 

\subsection{Analytic expression of $\ast$-ordered product}
\label{sec:*expr}
\indent

According to the differential expressions of $L_n^\pm$ and $T_n^{(k)}$ 
given by \eqref{Ln_diff} and \eqref{That}, we have the following operational relation
\beq
(q^{-z\partial} X_n) = q^{-n} X_n\,, \quad\quad 
X_n\in\mathscr{M}_T\{\hat{L}_n^\pm,\, \hat{T}_n^{(k)}\} \,. \label{*rule}
\eeq
Defining the analytic $\ast$-bracket expression similar to \eqref{f*g} for the element 
of $CZ^\ast$ algebra \eqref{CZCZ} $L_n^\pm \in\mathscr{M}_T$ as
\beq
(L_n^\eps L_m^\eta)_\ast =q^{\eps^{ab} z_a \partial_b} L_n^\eps L_m^\eta \,,
\label{moyal}
\eeq
where
\beq
(z_1,z_2)=(\eps z,\eta z) \,,\quad 
\eps^{ab}z_a\partial_b = z_1\partial_2 - z_2\partial_1 \,, \label{z1z2}
\eeq
we can reproduce the $\ast$-bracket defined in \eqref{CZ}
\beq
(L_n^\eps L_m^\eta)_\ast =q^{\eps m-\eta n}L_n^\eps L_m^\eta\,.
\eeq
By using this Moyal-like definition of the $\ast$-bracket, we can represent 
the $CZ^\ast$ algebra \eqref{CZCZ}. In the cases of other than $CZ^\ast$, 
${\mathcal{CZ}}^\ast$, $W^\ast_\infty$, $\tilde{W}^\ast_\infty$ as presented below, 
the detailed expression of the $\ast$-bracket is different for each algebra 
depending on the phase structure of the phase-shifted commutator, 
but the general form has the same structure as \eqref{moyal}.

The first concern is the $W_\infty^\pm$ algebra \eqref{WpmCom}. 
Since the size of its shifted phase in the bracket is in half of $CZ^\ast$, 
we can immediately define the analytic $\ast$-bracket for the element 
$W_n^{\eps(k)} \in\mathscr{M}_W$ as
\beq
(\Wp{n}{k}{\eps}\Wp{m}{l}{\eta})_\ast=
q^{\frac{1}{2}\eps^{ab} z_a \partial_b} \Wp{n}{k}{\eps} \Wp{m}{l}{\eta} \,, \label{Wmoyal}
\eeq
where $(z_1,z_2)$ is given by \eqref{z1z2}, and the same rule as \eqref{*rule} applies 
regarding $\Wp{n}{k}{\eps}$ as $X_n$. As a result, we obtain the $\ast$-bracket 
\beq
(\Wp{n}{k}{\eps}\Wp{m}{l}{\eta})_\ast=q^{\frac{\eps m-\eta n}{2}}
\Wp{n}{k}{\eps}\Wp{m}{l}{\eta}\,.
\eeq
Thus we verify that the $W_\infty^\ast$ algebra \eqref{Wpm*} can be written in 
terms of the $\ast$-bracket of $\mathscr{M}_W$ as
\beq
[W_n^{\eps (k)}, W_m^{\eta (l)}]_\ast =
[\frac{n(l-1)-\eps\eta m(k-1)}{2}] \sum_{\nu=\pm}
 C^\nu_{\eps,\eta} W_{n+m}^{\nu(\eps\nu k +\eta\nu l -2\eps\eta)} \,.
\eeq

{}~For the second kind of $\tilde{W}_\infty^\pm$ algebra \eqref{tildeWCom}, 
the bracket phase deviates from its first kind $W_\infty^\ast$, 
and we thus define 
\beq
(\tilde{W}_{n}^{\eps(k)}\tilde{W}_{m}^{\eta(l)})_\ast=
q^{\frac{1}{2}\eps^{ab} (z_a \partial_b-k_b)} 
\tilde{W}_{n}^{\eps(k)}\tilde{W}_{m}^{\eta(l)} \,, \label{wtilde*}
\eeq
where
\beq
(k_1,k_2)=(\eps k, \eta l)\,,\quad (z_1,z_2)=(\eps z, \eta z)\,. \label{k1k2}
\eeq
As shall be explained later, the phase shift $q^{-\frac{1}{2}\eps^{ab}k_b}$ is 
caused by the orthogonal transformation ($n-k$,$n+k$) of the phase 
coordinates$(n,k)$ suggested in \eqref{qplane_tr}. 
Consequently, by using this fact and in the same way as above, we derive 
the $\ast$-bracket
\beq
(\tilde{W}_{n}^{\eps(k)}\tilde{W}_{m}^{\eta(l)})_\ast=
q^{\frac{(\eps m-\eta l)-(\eta n-\eps k)}{2}} 
\tilde{W}_{n}^{\eps(k)}\tilde{W}_{m}^{\eta(l)} \,,
\eeq
and understand that the $\tilde{W}_\infty^\ast$ algebra \eqref{tildeW*} can be 
organized by the $\ast$-bracket of $\mathscr{M}_{\tilde{W}}$ as
\beq
[\tilde{W}_n^{\eps(k)}, \tilde{W}_m^{\eta(l)}]_\ast =
 [\frac{(n+1)(l+1)-\eps\eta(m+1)(k+1)+\eps\eta-1 }{2}] \sum_{\nu=\pm} 
C^\nu_{\eps,\eta} \tilde{W}_{n+m}^{\nu(\eps\nu k +\eta\nu l )} \,. 
\eeq

{}~Finally, the third example is the $\ast$-bracket \eqref{ML*} in $\mathscr{M}_L$
as to the ${\mathcal{CZ}}^\ast$ algebra. Defining the analytic relation 
\beq
(\Lp{n}{k}{\eps} \Lp{m}{l}{\eta})_\ast=
q^{-\frac{1}{2}\eps^{ab} z_a \partial_b} \Lp{n}{k}{\eps} \Lp{m}{l}{\eta}\,, \label{lnk*}
\eeq
and applying \eqref{*rule} with regarding $\Lp{n}{k}{\eps}$ as $X_n$ and 
\beq
(z_1,z_2)=(\eps k z, \eta l z)\,,  \label{mathCZz1z2}
\eeq
we identify that the $\ast$-bracket is given by
\beq
(L_n^{\eps(k)} L_m^{\eta(l)})_\ast =q^{\frac{\eta nl-\eps mk}{2} } L_n^{\eps(k)} L_m^{\eta(l)} \,,\quad  
\eeq
and reproduce the ${\mathcal{CZ}}^\ast$ algebra \eqref{tLtL}.

As can be seen from the above analytical expressions for the $\ast$-ordered product, 
all of them share the same structure $\eps^{ab}z_a\partial_b$, 
except for the phase shift $\eps^{ab}k_b$ in \eqref{wtilde*}. 
The only difference lies in the coefficient of the antisymmetric tensor $\eps^{ab}$. 

Although this may be obvious from the analysis of the MT operator representation, 
it is interesting to note that the shared structure is maintained while indicating 
the subtle differences in the coefficients and the phase shift, 
which determine the algebra-specific phase factors. 
The next section demonstrates that the phase shift arises from the phase space coordinate transformation in \eqref{qplane_tr}.

\subsection{Reduction from the Moyal bracket}
\label{sec:*moyal}
\indent

This section explores the connection between our $\ast$-bracket and the conventional Moyal bracket. It begins by elucidating the Moyal $\ast$-product of MT operators, drawing an analogy from the $T^2$ base in \eqref{txp}. The MT operator takes the following form after performing the coordinate transformation to the unit circle $z=e^{i\theta}$ in \eqref{Tmtheta}:
\beq
t_n^{(k)}=\exp{\frac{i}{\hbar}(nl_B\Theta_1 + k\frac{a^2}{l_B}\Theta_2)}\,.
\label{angleMT}
\eeq
Here, we normalize the dimensions of $\Theta_1$ and $\Theta_2$ so that 
$\mathrm{dim}(\theta_1,\theta_2)=(1,\hbar)$ corresponds to 
$\mathrm{dim}(x,p)=(1,\hbar)$, introducing 
\beq
\theta_1=\frac{l_B}{\hbar}\Theta_1=\theta\,,\quad
\theta_2=\frac{a^2}{l_B}\Theta_2=i\hbar(\frac{a}{l_B})^2(\partial_\theta+i\Delta)\,.
\eeq
The second commutation relation in \eqref{TTcom} is now
\beq
[\theta_i,\theta_j]=-i\hbar(\frac{a}{l_B})^2\eps^{ij}\,, \label{thithj}
\eeq
and the MT operator \eqref{angleMT} becomes
\beq
t_n^{(k)}=\exp{i(n\theta_1 + \frac{k}{\hbar}\theta_2)}
=\exp{i(n\theta_1 + k'\theta_2)}\,,  \label{tnk_theta}
\eeq
where $k'=k/\hbar$ is introduced with the dimension $\mathrm{dim}[k']=\hbar^{-1}$. 
If we treat $\theta_2$ as dimensionless, $k'$ should also be dimensionless. 
Similarly, in the case of $T^2$ base $\tau_{n,k}$ given in \eqref{txp} (or \eqref{T2base}) 
for the Moyal sine algebra \eqref{Msine}, 
we can also explicitly show the dimension dependence of $p$ so that the commutation 
relation $[x,p]=i\hbar$ is preserved in the parallel way to \eqref{tnk_theta} 
\beq
\tau_{n,k}=\exp{i(nx + \frac{k}{\hbar}p)}=\exp{i(nx + k'p)}\,. \label{taunk}
\eeq
Noticing the formal identity between \eqref{tnk_theta} and \eqref{taunk}, 
and focusing on \eqref{thithj}, we understand that the replacement
\beq
\hbar \quad\ra\quad -\hbar(\frac{a}{l_B})^2  \label{hbar2}
\eeq
can lead to the Moyal $\ast$-product for the MT operator base \eqref{tnk_theta}. 
The coefficient part of $\hbar$ in \eqref{hbar2} can be absorbed in the definition 
of $\theta^{ab}$ as
\beq
\theta^{ab}=-\theta \eps^{ab}\quad \ra\quad(\frac{a}{l_B})^2\eps^{ab}\,, \label{thab}
\eeq
yielding the replacement in the Moyal product operator
\beq
e^{\frac{i\hbar}{2}\theta^{ab}\partial_1^a\partial_2^b}\quad\ra\quad
\exp{{\frac{i\hbar}{2}(\frac{a}{l_B})^2\eps^{ab}\partial_1^a\partial_2^b}} \,.
\label{MT*}
\eeq
The Moyal $\ast$-product for the MT operator is therefore expressed as
\beq
t_n^{(k)}\ast t_m^{(l)}=e^{\frac{i\hbar}{2}\theta^{ab}\partial_1^a\partial_2^b}
t_n^{(k)}t_m^{(l)}
=\exp{{\frac{i\hbar}{2}(\frac{a}{l_B})^2\eps^{ab}\partial_1^a\partial_2^b}}
t_n^{(k)}t_m^{(l)}\,.  \label{t*t}
\eeq

The change in the fundamental commutation relation 
from $[x,p]=i\hbar$ to \eqref{thithj} also affects the definition of $q$. 
Treating $k'$ as dimensionless implies that $p$ and $\theta_2$ must also be dimensionless. Consequently, $\theta^{ab}$, which was dimensionless in \eqref{thab}, acquires the dimension of $k'$. 
In other words, $\mathrm{dim}[\hbar\theta^{ab}]=\mathrm{dim}[\hbar\theta]=1$. 
This observation suggests that $\theta$ represents the magnitude of quantum space fluctuations with the unit of $\hbar$, and hence $\theta$ might be termed the {\it quantum dimension}. 
With this dimensionless formulation, $q$ is expressed in terms of the quantum dimension $\theta$ as:
\beq
q=e^{i\hbar\theta} \quad\ra\quad e^{i(\frac{a}{l_B})^2}\,. \label{qdim}
\eeq

Hereafter, we explain how to derive the $\ast$-product expression for each case presented in Section~\ref{sec:*expr} from the Moyal $\ast$-product \eqref{t*t}. Remembering the eigenvalue expression for the Moyal $\ast$-product 
of $T^2$ base
\beq
\tau_{n,k}\ast \tau_{m,l}=e^{\frac{i\hbar}{2}\theta^{ab}\partial_1^a\partial_2^b}
\tau_{n,k}\tau_{m,l}=e^{\frac{i}{2}\theta(nl-mk)}\tau_{n,k}\tau_{m,l} \label{tau*prod}
\eeq
with the forward/backward derivative action of $(x,p)$ on the base
\beq
(\partial_1^x,\partial_2^x)\ra (in,im)\,,\quad (\partial_1^p,\partial_2^p)\ra (ik',il')\,,
\eeq
the derivative action of $(\theta_1,\theta_2)$ on the MT base \eqref{tnk_theta} 
can be estimated by taking account of eigenvalues on $t_n^{(\eps k)}t_m^{(\eta l)}$ 
in the similar way as \eqref{tau*prod}:
\beq
(\partial_1^{\theta_1},\partial_2^{\theta_1})\ra (in,im)\,,\quad
 (\partial_1^{\theta_2},\partial_2^{\theta_2})\ra (i\eps k',i\eta l')\,. \label{diffeigen}
\eeq
{}~Furthermore, paying attention to the relation $\partial_\theta=iz\partial_z$, 
we can transform the $\theta_1$ derivatives into the $z$ derivatives
\beq
(\partial_1^{\theta_1},\partial_2^{\theta_1})\ra (iz\partial_1^z,iz\partial_2^z)\,,\quad
 (\partial_1^{\theta_2},\partial_2^{\theta_2})\ra (i\eps k',i\eta l')\,. \label{repdiff}
\eeq
As mentioned in Section~\ref{sec:*expr}, we may introduce a proper 
coefficient $\alpha$ on the antisymmetric tensor $\eps^{ab}$ 
in accord with an algebra, thus generalizing \eqref{thab} as 
\beq
\theta^{ab}=-\theta\eps^{ab}\,,\quad \theta=-\frac{\alpha}{\hbar}(\frac{a}{l_B})^2\,.
\label{theta2}
\eeq
Note that $(k',l')=(k,l)$ since we have set $\mathrm{dim}[\hbar\theta]=1$. 
As a consequence of \eqref{repdiff} and \eqref{theta2}, the Moyal operator 
\eqref{MT*} yields as follows: 
\begin{align}
e^{\frac{i\hbar}{2}\theta^{ab}\partial_1^a\partial_2^b}
&=\exp{\frac{i}{2}\alpha(\frac{a}{l_B})^2(\partial_1^{\theta_1}\partial_2^{\theta_2}
-\partial_1^{\theta_2}\partial_2^{\theta_1} )}  \nn \\
&\ra\exp{\frac{i}{2}\alpha(\frac{a}{l_B})^2(\eps kz\partial_2^z-\eta lz\partial_1^z)}\,.
\label{m-factor}
\end{align}

When converting to the $\ast$-ordered product, there is one remaining point to note. 
While setting $\alpha=-1$ reproduces the $\ast$-ordered product \eqref{lnk*} of 
$\mathcal{CZ}^\ast$, as evident from substituting \eqref{mathCZz1z2} into 
\eqref{m-factor}, a slight adjustment is necessary for other algebras such as $CZ^\ast$,  $W_\infty^\ast$, and $\tilde{W}_\infty^\ast$.

To make this adjustment, we recall that in the $\ast$-bracket factor of $\mathscr{M}_T$,
we had to set $k=l=\pm2$ as the weight for the $CZ^\pm$ operators. 
To reproduce the $\ast$-bracket expressions for $CZ^\ast, W_\infty^\ast$ 
and $\tilde{W}_\infty^\ast$, we need to incorporate this kind of final operation 
into \eqref{m-factor}. The procedure involves assigning unique values for each algebra to the weight-related parts, $\eps k$ and $\eta l$. 
In the case of $CZ^\ast$, 
setting the weight to $l=k=2$ reproduces the $\ast$-ordered product \eqref{moyal} 
for the choice $\alpha=1$ and the coordinates \eqref{z1z2}. 
In the case of $W_\infty^\ast$, the weight to $k=l=1$ similarly reproduces the 
$\ast$-ordered product \eqref{Wmoyal} of $\mathscr{M}_W$ 
for the same $\alpha$ and coordinates as those of $CZ^\ast$.

In the case of $\tilde{W}_\infty^\ast$, 
it is clear at a glance that the $\ast$-ordered product cannot be reproduced 
simply by the same procedure, because it differs by a factor of 
$-\frac{1}{2}\eps^{ab}k_b$ compared to $W_\infty^\ast$. 
Nevertheless, the $\ast$-ordered product for $\tilde{W}_\infty^\ast$ can be derived by rearranging the phase space coordinate system and following a similar procedure, as explained below.

In the first place, changing the coordinates from $(\theta_1,\theta_2)$ to 
$(\theta_+,\theta_-)$, where $\theta_\pm=\frac{1}{2}(\theta_1\pm\theta_2)$, 
let us introduce
\beq
\partial_a^{\theta_\pm}=\partial_a^{\theta_1}\pm\partial_a^{\theta_2}\,,
\label{newtheta}
\eeq
which induces the eigenvalue transformation \eqref{qplane_tr}. 
If we rewrite the operator part of Moyal $\ast$-product as
\beq
\eps^{ab}\partial_a^{\theta_1}\partial_b^{\theta_2}=\frac{1}{2}
\eps^{ab}\partial_a^{\theta_+}\partial_b^{\theta_-} \,,
\eeq
we then derive the following eigenvalue expressions instead of 
\eqref{diffeigen} and \eqref{repdiff}, using the notation \eqref{k1k2}
\begin{align}
&(\partial_1^{\theta_-},\partial_2^{\theta_-})\,\quad\ra\quad(in-ik_1,im-ik_2) \label{nkmix}\\
&(\partial_1^{\theta_+},\partial_2^{\theta_+})\,,\quad\ra\quad(iz_1\partial_1^z+ik_1,iz_2\partial_2^z+ik_2) \,.
\end{align}
Remembering the eigenvalue relations
\beq
m\partial_1^{\theta_1}-n\partial_2^{\theta_1}=0\,,\quad
k_1\partial_2^{\theta_2}-k_2\partial_1^{\theta_2}=0\,,
\eeq
we can alternate \eqref{m-factor} to obtain
\beq
e^{\frac{i\hbar}{2}\theta^{ab}\partial_1^a\partial_2^b} \ra
\exp{\frac{i}{4}\alpha(\frac{a}{l_B})^2(kz_1\partial_2^z-lz_2\partial_1^z+mk_1-nk_2)}\,.
\label{newfactor}
\eeq
{}~Finally, choosing $\alpha=-1$, and setting $n,m,k,l=2$ as the weight on RHS of 
\eqref{newfactor}, we derive the $\ast$-bracket operator of 
$\mathscr{M}_{\tilde{W}}$ appeared in \eqref{wtilde*}
\beq
q^{\frac{1}{2}\eps^{ab}(z_a\partial_b^z-k_b)}\,.
\eeq
We should note that it is not unnatural to set $(k,l)$ together with $(n,m)$, 
because the coordinate transformation \eqref{newtheta} mixes the two as 
seen in \eqref{nkmix} and cannot be handled separately.

We have shown that $\mathcal{CZ}^\ast$ exhibits a one-to-one correspondence with the Moyal product operator without fixing the weight. In contrast, $CZ^\ast, W_\infty^\ast$, and $\tilde{W}_\infty^\ast$ require weight fixing for the reduction. Until now, we have used the term "weight" without assigning a specific meaning. However, considering its role as a $\theta$ multiplier that contributes to the unique structure constant of the $\ast$-product operator for each algebra, it would be appropriate to refer to it as the {\it quantum dimensional weight}.

\setcounter{equation}{0}
\section{Conclusions and Outlook}
\indent

In this paper, we have investigated the generalizations of the CZ algebra and explored their algebraic properties and their relationship to physical models. In each case, we uncovered intriguing properties and identified several open questions for further exploration.

In Section~\ref{sec:GCZ}, 
we considered the $\mathcal{CZ}$ algebras as a generalization of $CZ$ algebras. 
We defined new generators $L_n^{\pm(k)}$ by introducing a scaling operation to 
$CZ^\pm$ generators. We presented the most general algebraic relations 
with external phase factors and investigated the mathematical properties of 
$L_n^{\pm(k)}$, such as their $\ast$-bracket representation and $CZ$ subalgebras.
In Section~\ref{sec:winf}, 
we decomposed $L_n^{\pm(k)}$ into two types of Hom-Lie type $W_\infty$ algebras 
and studied the $\ast$-bracket structure of each. 
In Section~\ref{sec:XYrep}, 
we obtained the matrix representations of these algebras and derived 
the Hamiltonian $H_g$ realized by $\mathcal{CZ}^-$ based on TBM 
in terms of physical relevance. 
In Section~\ref{sec:*CZW}, 
we discussed the relationship between the $\ast$-bracket structures of 
these algebras and the Moyal $\ast$-product of MT operators. 

Through our study of the generalized CZ algebras, we have identified four directions of inquiry, outlined below.
While each direction has yielded intriguing findings, it is evident that we have only glimpsed a portion of the overarching coherent picture. Significant work remains to develop a comprehensive understanding of these research avenues.

The first direction focuses on the structural properties of the minimal subalgebras and the $\ast$-bracket. In Section~\ref{sec:GCZ*alg}, 
we introduced new algebras $\mathcal{CZ}^\pm$ and $\mathcal{CZ}^\ast$ expressed in terms of a new $\ast$-bracket of $\mathscr{M}_L$ by choosing a specific external phase factor. While the weight of the $\ast$-bracket of $\mathscr{M}_T$ was set to 2 for the $CZ^\ast$ generators, the weight naturally corresponds to $k$, the upper index on $L_n^{\pm(k)}$, for the $\ast$-bracket of $\mathscr{M}_L$. 
In Section~\ref{sec:subAlg}, we showed that the minimal subalgebra of the algebra 
$\mathcal{CZ}^\ast$ satisfied by $L_n^{\pm(k)}$ is $CZ^\ast$, and we showed 
that $CZ^\ast$ and $\mathcal{CZ}^\ast$ have the same $\ast$-bracket structure. 
However, since the treatment of weight differs (it is 2 for $CZ^\ast$, 
but taken as 0 for $\mathcal{CZ}^\ast$), the structure constants inevitably differ 
due to the difference in weight. Nevertheless, by reconstructing the whole 
based on the $k=2$ subalgebra with $L_n^{\pm(2)}$, 
there remains a possibility of resolving this structural issue.

The second direction concerns the functional features of the $L_n^{\pm(k)}$ 
generators. While exhibiting some nontrivial properties, such as containing $CZ$ as a subalgebra and being a scaled version of $CZ$, these generators manifest as a mere sum of two MT operators.  As examined in Section ~\ref{sec:winf}, these two MT operators provide two deformation types for $w_{1+\infty}$. 
The generators of $CZ^\ast$ consist of $T_n^{(0)}$ and $T_n^{(\pm2)}$, 
and are constructed as a linear combination of the commutative $CZ$ representation $\VT{n}{0}$ and the non-commutative but $\ast$-bracket commutative 
one $\VT{n}{\pm2}$, exhibiting a simple structure~\cite{AS2}. 
On the other hand, the generators of $L_n^{\pm(k)}$ consist of a combination of 
$W_n^{\pm(k)}$ and $\tilde{W}n^{\pm(k)}$, corresponding to the deformations $W_\infty^\ast$ and $\tilde{W}_\infty^\ast$ of $w_{1+\infty}$ described in \eqref{winf1} and \eqref{winf2}, respectively. The part corresponding to the commutative $CZ$ representation $\VT{n}{0}$, either $W_n^{\pm(2)}$ or $\tilde{W}_n^{\pm(0)}$, 
is contained in $W_\infty^\ast$ and $\tilde{W}_\infty^\ast$ as the respective subalgebras \eqref{CZsub1} and \eqref{CZsub2}. This structure is quite different from $CZ^\ast$, and whether there is a more organized understanding remains an open question.

As a third avenue, while deriving interesting results concerning physical models, 
we have only found fragmentary achievements, and the issue of parameter selection 
has emerged to obtain a solution with an overall perspective. 
In Section~\ref{sec:XYrep}, we obtained matrix representations of $\mathcal{CZ}^\ast$,
$W_\infty^\pm$ and others, and derived their relationship with the Hamiltonian family 
of the TBM models. For $W_\infty^\pm$, as suggested by \eqref{TBHnk}, $H_{(n,k)}$ 
can be written in the form of a trace sum along a closed loop of the translation operator 
on the quantum line, interpreted from \eqref{tau_decomp} and \eqref{qplane_tr}, 
resulting in a more streamlined form compared to the $CZ^\ast$ case~\cite{AS2}. 
Placing a central core basic Hamiltonian $H_0$ at the site $(n,k)$, 
and expressing the overall fluctuation effect across the quantum plane 
as a superposition of peripheral interactions $H_{(n,k+l)}$, 
we can represent the total Hamiltonian, for example, as:
\beq
H=H_0 + \sum_l  g_l H_{(n,k+l)}\,. \label{Htotal}
\eeq
By choosing an appropriate term for $H_0$, for example $H'_{(n,k)}$,  and 
examining only the $l=-2$ effect, $H$ becomes $H'_g$ given by \eqref{Hg'}, and 
can be represented by the $\mathcal{CZ}^-$ generators 
$\mathrm{L}_n^{-(k)}$ \eqref{vki}, $\mathscr{L}_n^{-(k)}$ \eqref{vk+} 
and $\mathscr{K}_n^{-(k)}$ \eqref{vk-} depending on the value of $g_l=g$. 
Moreover, applying the phase rotation $\mathbb{L}_n^{-(k)}$ to the case \eqref{vki}, 
$H'_g$ with an imaginary coupling $i$ transforms into $H_g$ 
with a real coupling $g_l=-1$ to another central Hamiltonian $H_0=H_{(n,k)}$.

Particularly, for the case $k=0$, we have obtained $H=H_{(\pm n, -2)}$ represented 
by the $CZ^-$ generators \eqref{CZtL} and \eqref{CZKn} for $g_l=\pm1$. 
However, since the previous paper discussed the relationship with $H_{(n,2)}$ 
through \eqref{HgL} and \eqref{CZtL}, the discussion is not yet fully consistent. 
As a possibility, $H=H_{(n,2)}$ may appear as an effect from different values of $l$, 
or it may emerge from a different $H$ than \eqref{Htotal} in connection to the $k=2$ 
subalgebra mentioned earlier. Furthermore, to achieve overall consistency, 
we may need to consider redefining the sign of $k$, as required by \eqref{ML*2}. 
Since there are still degrees of freedom remaining for overall adjustment, 
such as $k$ and $g$, the possibility of representing with $\mathcal{CZ}^+$ 
or deriving $H_{(n,2)}$ is still open.

As for the fourth perspective, there are interesting findings on the Moyal 
product. The two components $\tilde{W}_n^{\pm(k)}$ and $W_n^{\pm(k)}$ obtained by 
linearly decomposing $L_n^{\pm(k)}$ satisfy $\tilde{W}_\infty^\pm$ and $W_\infty^\pm$ 
respectively. Furthermore, including the $\pm$ interactions, they can be collectively 
organized into the prescribed $\ast$-bracket forms like $\tilde{W}_\infty^\ast$ and 
$W_\infty^\ast$, as seen in Section~\ref{sec:W*}. 
Although these forms resulted from artificially choosing the external factors to satisfy 
\eqref{WpmCom2} and \eqref{tildeWCom2}, a certain objectivity has been ensured 
in Section~\ref{sec:*CZW} for the common forms of $\mathcal{CZ}^\ast$, 
$W_\infty^\ast$, and $\tilde{W}_\infty^\ast$ by verifying their analytical expressions 
for the $\ast$-bracket through the reduction prescription from the Moyal 
$\ast$-product. 

In particular, the $\ast$-bracket of $\mathcal{CZ}^\ast$ corresponds equivalently 
to the Moyal $\ast$-product, and the other algebras are reproduced as 
projections from the Moyal $\ast$-product via weight reduction 
(see Section~\ref{sec:*moyal}). 
Unlike $W_\infty^\ast$, $\tilde{W}_\infty^\ast$ requires the phase transformation 
\eqref{qplane_tr}, and this difference can serve as a reason to treat 
$\tilde{W}_\infty^\ast$ and $W_\infty^\ast$ independently, reinforcing the property 
of $T_n^{(0)}$ and $T_n^{(\pm2)}$ behaving independently in $CZ^\pm$~\cite{AS2}. 
However, the weight reduction prescription is artificial, and the mechanism 
by which it is brought about remains obscure.

{}~Finally, we explain the origin of naming it the "{\it quantum dimension}" for 
$\theta$. For this, we focus on the relationship between the classical and the quantum Moyal form. Writing the commutation relation $[x,p]=i\hbar$ as
\beq
[\hat{x}_a,\hat{x}_b]=i\delta\,,\quad\,\,\delta=\hbar\,,
\eeq
then \eqref{thithj} can be written in the same form with $\delta=-\hbar(\frac{a}{l_B})^2$. 
In classical theory, the Moyal product takes the form
\beq
e^{i(nx_1+kx_2)}\ast e^{i(mx_1+lx_2)}=
e^{i\frac{\hbar\theta}{2}(nl-mk)}e^{i(n+m)x_1+i(k+l)x_2}\,,
\eeq
whereas in quantum theory, using the Campbell-Baker-Hausdorff formula, we have
\begin{align}
e^{i(n\hat{x}_1+k\hat{x}_2)}\ast e^{i(m\hat{x}_1+l\hat{x}_2)}
&= e^{i\frac{\hbar\theta}{2}(nl-mk)}e^{i(n\hat{x}_1+k\hat{x}_2)} e^{i(m\hat{x}_1+l\hat{x}_2)}\nn\\
&=e^{i\frac{\hbar\theta}{2}(nl-mk)}e^{-i\frac{\delta}{2}(nl-mk)}
e^{i(n+m)\hat{x}_1+i(k+l)\hat{x}_2}\,.
\end{align}

Hence, in quantum theory, by redefining $\theta$ as
\beq
\hbar\theta\,\quad\ra\quad \hbar\theta+\delta\,, \label{qhtheta}
\eeq
we can reduce to the same form as classical theory. 
Conversely, by redefining $\theta$ from classical theory, we can transition to 
the same form as quantum theory.

If we consider this transition width $\delta$ as the quantum fluctuation of $\theta$, 
then $q$ inherits the nature of quantum fluctuation through the relation 
$q=e^{i\hbar\theta}$ in the form of integer powers of $e^{i\delta}$. 
Since we can transition to the classical formula any number of times through \eqref{qhtheta} with each application of the Moyal $\ast$-operation, $q^k$ can be interpreted as a parameter characterizing the quantum space with 
the number $k$ of transitions between classical and quantum theories. 
That is the reason we called $\theta$ the "{\it quantum dimension}" in \eqref{qdim}.

To be more precise, let's consider the case of $\mathcal{CZ}^\ast$, for example. 
Taking $\alpha=-1$ in \eqref{theta2}, we have
\beq
\hbar\theta=-\alpha(\frac{a}{l_B})^2=(\frac{a}{l_B})^2\,.
\eeq
In the dimensionless system $\mathrm{dim}[\hbar\theta]=1$, we should 
divide the $\delta$ in the above commutation relation by $\hbar$. 
Observing the change in $\hbar\theta$ due to a single fluctuation, we have
\beq
\hbar\theta\quad\ra\quad \hbar\theta+\frac{\delta}{\hbar} 
=\hbar\theta-(\frac{a}{l_B})^2 =0\,.
\eeq
Denoting the number of fluctuations as $1-k$, similarly we have
\beq
\hbar\theta\quad\ra\quad k\hbar\theta\,,
\eeq
and the change in $q$ appears in the exponent of $q$ as
\beq
q=e^{i\hbar\theta}\quad\ra\quad q^k\,.
\eeq

In this paper, we have derived the matrix representations of the generalized CZ algebras and discussed their relationship to the Hamiltonian of TBM. This result indicates a possible use of $q$-deformation in condensed matter physics and suggests that $q$-deformed algebras offer a useful mathematical tool for analyzing physical systems.

We explored various aspects of $\mathcal{CZ}^\ast$ in this paper,  including the interplay between $k=2$ subalgebras and $CZ^\ast$, the tight-binding model Hamiltonian, and the Moyal $\ast$-product structures.  We anticipate that this comprehensive investigation will shed light on unresolved questions surrounding $\mathcal{CZ}^\ast$ and $CZ^\ast$~\cite{CKL}-\cite{super3}. By doing so, we hope to unveil profound connections and novel insights into related domains, such as the $q$-deformation of thermodynamics and quantum mechanics~\cite{TS}-\cite{WSC3}, thereby potentially paving the way for future research directions.

Furthermore, regarding the deformation and generalization of Lie algebras, we find it intriguing to explore the direction of quasi-Hom-Lie (QHL) algebras. Ref. \cite{Hom2} defined a QHL-algebra by twisting maps for Jacobi identity and skew-symmetry, proposing a generalization of Hom-Lie algebras. The $CZ^\ast$ algebras that we investigated has a similar form to the $q$-Virasoro algebra discussed in \cite{HPJ}, and these might be regarded as specific cases of QHL-algebras. Ref. \cite{HPJ} suggests a potential connection of the $q$-Virasoro algebra to the quantization of matrix models, as open string algebras and $q$-Virasoro algebras can be understood as part of the theoretical framework of matrix models. Investigating concrete aspects along this line might be an interesting pursuit.

\section*{CRediT authorship contribution statement} 
 
\textbf{Haru-Tada Sato:} Writing-Original Draft, Conceptualization, Methodology, 
Investigation, Validation.

\section*{Declaration of competing interest}

The author declares that we have no known competing financial interests or personal relationships that could have appeared to influence the work reported in this paper.

\section*{Data availability}

No data was used for the research described in the article. 


\newcommand{\NP}[1]{{\it Nucl.{}~Phys.} {\bf #1}}
\newcommand{\PL}[1]{{\it Phys.{}~Lett.} {\bf #1}}
\newcommand{\Prep}[1]{{\it Phys.{}~Rep.} {\bf #1}}
\newcommand{\PR}[1]{{\it Phys.{}~Rev.} {\bf #1}}
\newcommand{\PRL}[1]{{\it Phys.{}~Rev.{}~Lett.} {\bf #1}}
\newcommand{\PTP}[1]{{\it Prog.{}~Theor.{}~Phys.} {\bf #1}}
\newcommand{\PTPS}[1]{{\it Prog.{}~Theor.{}~Phys.{}~Suppl.} {\bf #1}}
\newcommand{\MPL}[1]{{\it Mod.{}~Phys.{}~Lett.} {\bf #1}}
\newcommand{\IJMP}[1]{{\it Int.{}~Jour.{}~Mod.{}~Phys.} {\bf #1}}
\newcommand{\JPA}[1]{{\it J.{}~Phys.} {\bf A}:\ Math.~Gen. {\bf #1}~}
\newcommand{\JHEP}[1]{{\it J.{}~High Energy{}~Phys.} {\bf #1}}
\newcommand{\JMP}[1]{{\it J.{}~Math.{}~Phys.} {\bf #1} }
\newcommand{\CMP}[1]{{\it Commun.{}~Math.{}~Phys.} {\bf #1} }
\newcommand{\LMP}[1]{{\it Lett.{}~Math.{}~Phys.} {\bf #1} }
\newcommand{\doi}[2]{\,\href{#1}{#2}\,}  


\end{document}